\documentclass[aps,10pt,superscriptaddress,notitlepage,amsmath,amssymb,twocolumn,pra]{revtex4-2}

\usepackage{graphicx}
\usepackage{dcolumn}
\usepackage{bm}
\usepackage{enumitem}
\usepackage{amsthm}
\usepackage{color}
\usepackage[colorlinks=true,linkcolor=red,urlcolor=black,citecolor=blue]{hyperref}

\theoremstyle{definition}
\newtheorem{defi}{Definition}

\theoremstyle{plain}
\newtheorem{prop}{Proposition}

\theoremstyle{remark}
\newtheorem*{prf}{Proof}
\newtheorem{theo}{Theorem}

\begin{document}

\preprint{APS/123-QED}

\title{Continuous-variable quantum digital signatures that can withstand coherent attacks}

\author{Yi-Fan Zhang}\thanks{These authors contributed equally.}
\affiliation{National Laboratory of Solid State Microstructures and School of Physics, Collaborative Innovation Center of Advanced Microstructures, Nanjing University, Nanjing 210093, China}
\affiliation{Department of Physics and Beijing Key Laboratory of Opto-electronic Functional Materials
and Micro-nano Devices, Key Laboratory of Quantum State Construction and Manipulation (Ministry of Education),
Renmin University of China, Beijing 100872, China}
\author{Wen-Bo Liu}\thanks{These authors contributed equally.}
\affiliation{National Laboratory of Solid State Microstructures and School of Physics, Collaborative Innovation Center of Advanced Microstructures, Nanjing University, Nanjing 210093, China}
\affiliation{Department of Physics and Beijing Key Laboratory of Opto-electronic Functional Materials
and Micro-nano Devices, Key Laboratory of Quantum State Construction and Manipulation (Ministry of Education),
Renmin University of China, Beijing 100872, China}
\affiliation{College of Information and Communication, National University of Defense Technology, Wuhan 430010, China}
\author{Bing-Hong Li}
\affiliation{National Laboratory of Solid State Microstructures and School of Physics, Collaborative Innovation Center of Advanced Microstructures, Nanjing University, Nanjing 210093, China}
\affiliation{Department of Physics and Beijing Key Laboratory of Opto-electronic Functional Materials
and Micro-nano Devices, Key Laboratory of Quantum State Construction and Manipulation (Ministry of Education),
Renmin University of China, Beijing 100872, China}
\author{Hua-Lei Yin}\email{Contact author: hlyin@ruc.edu.cn}
\affiliation{Department of Physics and Beijing Key Laboratory of Opto-electronic Functional Materials
and Micro-nano Devices, Key Laboratory of Quantum State Construction and Manipulation (Ministry of Education),
Renmin University of China, Beijing 100872, China}
\affiliation{National Laboratory of Solid State Microstructures and School of Physics, Collaborative Innovation Center of Advanced Microstructures, Nanjing University, Nanjing 210093, China}
\affiliation{Beijing Academy of Quantum Information Sciences, Beijing 100193, China}
\author{Zeng-Bing Chen}\email{Contact author: zbchen@nju.edu.cn}
\affiliation{National Laboratory of Solid State Microstructures and School of Physics, Collaborative Innovation Center of Advanced Microstructures, Nanjing University, Nanjing 210093, China}

\date{\today}

\begin{abstract}
Quantum digital signatures (QDSs), which utilize correlated bit strings among sender and recipients, guarantee the authenticity, integrity, and nonrepudiation of classical messages based on quantum laws. Continuous-variable (CV) quantum protocol with heterodyne and homodyne measurement has obvious advantages of low-cost implementation and easy wavelength division multiplexing. However, security analyses in previous researches are limited to the proof against collective attacks in finite-size scenarios. Moreover, existing multibit CV QDS schemes have primarily focused on adapting single-bit protocols for simplicity of security proof, often sacrificing signature efficiency. Here, we introduce a CV QDS protocol designed to withstand general coherent attacks through the use of a cutting-edge fidelity test function, while achieving high signature efficiency by employing a refined one-time universal hashing signing technique. Our protocol is proved to be robust against finite-size effects and excess noise in quantum channels. In simulation, results demonstrate a significant reduction of eight orders of magnitude in signature length for a megabit message signing task compared with existing CV QDS protocols and this advantage expands as the message size grows. Our work offers a solution with enhanced security and efficiency, paving the way for large-scale deployment of CV QDSs in future quantum networks. 
\end{abstract}
\maketitle


\section{\label{sec1}Introduction}

Quantum digital signatures (QDSs) have emerged as a solution to ensure the authenticity, integrity, and nonrepudiation of messages~\cite{menezes1997handbook}, leveraging the principles of quantum mechanics to provide information-theoretic security~\cite{nielsen2002quantum,yin2022experimental}. In the literature, QDS schemes do not assume the existence of a broadcast channel or a trusted authority~\cite{amiri2015unconditionally}, which can be regarded as the basis requirement of designing QDS protocol. In 2001, Gottesman and Chuang introduced the first single-bit QDS scheme based on a quantum one-way function~\cite{gottesman2001quantum}, which is expected to be used to sign binary messages, i.e., yes or no. Despite the theoretical proof of security under the assumption of a secure quantum channel, the practical implementation of this scheme is hindered by the requirement of high-dimensional single-photon states and perfect quantum swap tests. Over the past decades, significant efforts have been made to develop practical single-bit QDS schemes by relaxing certain assumptions~\cite{andersson2006experimentally,clarke2012experimental,dunjko2014quantum}. A significant breakthrough came in 2016 with the proposal of two unconditionally secure single-bit QDS protocols, even with an insecure quantum channel~\cite{yin2016practical, amiri2016secure}, which push QDS to the experimental demonstration stage. Since then, numerous experimental demonstrations~\cite{collins2014realization,  yin2017experimentala, collins2017experimental, yin2017experimental, roberts2017experimental, zhang2018proof, roehsner2018quantum, an2019practical, ding2020280km,pelet2022unconditionally,chapman2024entanglement} and theoretical advancements~\cite{lu2021efficient,zhang2021twin,weng2021secure,qin2022quantum} have confirmed the feasibility of implementing single-bit QDSs in practice. 

Actually, there are two main categories of schemes for encoding and measuring quantum states, including discrete-variable (DV) and continuous-variable (CV) systems. The DV method, with a rich history dating back to early studies~\cite{brassard1984quantum, ekert1991quantum}, relies on photon-number detection with single photons and owns extensive studies on its theoretical security properties~\cite{shor2000simple, xie2022breaking}. However, practical implementation of the DV method encounters challenges, such as the need for delicate single-photon sources and detectors, prompting ongoing research efforts to address these obstacles~\cite{liu2019experimental, pittaluga2021600km,gu2022experimental, wang2022twin, clivati2022coherent, zhou2023experimental}. In contrast, the CV method, which encodes information into continuous degrees of freedom, such as the phase of the electromagnetic field, utilizes coherent optical setups for homodyne or heterodyne measurements~\cite{ralph1999continuous, grosshans2002continuous}. Compared with the DV scheme, the CV quantum protocol offers more efficient, high-rate, and cost-effective procedures for preparing and measuring quantum states, making it more compatible with existing large-scale communication network infrastructure~\cite{huang2015continuous, kumar2015coexistence, huang2016field, karinou2017experimental, liu2021homodyne, karinou2018integration, eriksson2018coexistence, eriksson2019wavelength, eriksson2020wavelength,jain2022practical, hajomer2024long}.

In recent years, several CV QDS protocols~\cite{croal2016freespace, thornton2019continuous,richter2021agile,zhao2021quantum,zhao2023quantum} have been proposed and experimentally demonstrated for signing one-bit message. However, the security of these solutions against general coherent attacks in finite-size regime with quantified failure probabilities remains unproven. The need for theoretical advancement in security proof becomes a major obstacle towards the wide application of CV QDS. A promising route in direction can be adopting some DV techniques in security analysis~\cite{matsuura2021finite}.

In addition to that, extending to multibit signature is another challenging task, as simple concatenation without careful coding of single-bit signature can be vulnerable to attacks~\cite{wang2015security}. To tackle this issue, various approaches have been proposed, primarily involving the encoding of raw messages into new strings and the iterative application of the single-bit protocol to each bit~\cite{wang2015security, wang2017postprocessing, zhang2019high, cai2019cryptanalysis,zhao2021multibit}. Reference~\cite{wang2015security} encodes each message bit 0(1) into 000(010) and appends a symbol word 111 to both the head and tail of the message. Reference~\cite{zhang2019high} follows a specific coding rule that inserts a bit 0 every $x$ message bits. Reference~\cite{zhao2021multibit} signs each message bit with increasing signature lengths without extra coding. Obviously, these methods sacrifice the signature efficiency in exchange for security and thus their achieved signature rates are low and insufficient for practical implementation.

Fortunately, one-time universal hashing (OTUH) QDSs~\cite{yin2022experimental} have emerged as a solution to the multibit message signing with very high efficiency and unconditional security, containing two key attributes. On the one hand, the users' secret sharing relationship establishes a strict asymmetry between the signer and the recipient. In this regard, once two recipients collaborate, they will possess identical information (keys), thereby preventing the repudiation attempts by the signer. On the other hand, the use of universal hashing in each update guarantees a strict one-way characteristic, rendering it impossible for the recipient to infer the message from the encrypted hash value. Notably, even with imperfect discrete-variable quantum state~\cite{li2023one}, where an attacker obtains part of the information, OTUH QDSs remain unconditionally secure due to the inherent compression properties of the universal hash functions. Additionally, OTUH QDSs with a discrete-variable system has been experimentally validated and used to build other quantum protocols~\cite{cao2024experimental,weng2023beating,jing2024experimental}, such as quantum e-commerce and quantum Byzantine consensus.

\begin{figure*}
    \centering
    \includegraphics[width=\linewidth]{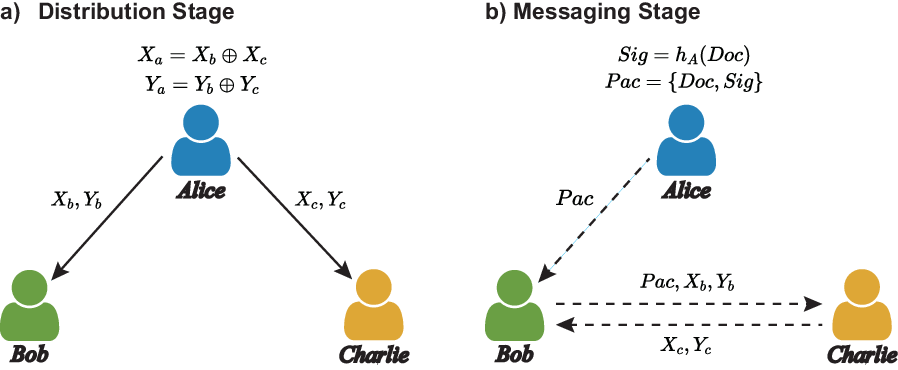}
    \caption{The schematic of the CV QDS protocol. Solid (Dashed) lines represents quantum (classical) channels. The protocol consist of two main parts: distribution stage and messaging stage. (a) In the distribution stage, Alice sends optical pulses through quantum channels to Bob and Charlie to generate shared keys in the distribution stage. (b) In the messaging stage, Alice generates the signature through OTUH and sends the packet containing the message and signature to Bob. Bob then sends his keys and received information to Charlie, who later sends his keys to Bob. Bob and Charlie use their own and received keys to infer Alice's keys and then perform OTUH tests to verify the signature.}
    \label{fig1}
\end{figure*}

In this paper, we present a CV QDS protocol based on the OTUH method~\cite{yin2022experimental,li2023one}. As an efficient method which directly signs the hash value of multibit messages with just one key string, its application significantly improves the signature efficiency. In addition to that, we adopt discrete-modulated CV approach inspired by a state-of-the-art quantum key distribution work~\cite{matsuura2021finite} for the generation of shared keys, which guarantees security against general coherent attacks in finite-size regime for the first time. Thanks to the seamlessly integration of both parts, we provide the complete security proof of multibit CV QDS scheme. In summary, our protocol efficiently signs multibit messages with information-theoretical security, achieving higher signature rates in short distances and requiring inexpensive and readily available apparatus. To demonstrate the superiority of our protocol in performance, we also conduct a numerical simulation under different conditions. From the results, we would find that our protocol can outperform previous CV QDS protocols by over eight orders of magnitude in terms of signature rates for a megabit message signing task over a 25km fiber between the sender and recipient. Furthermore, our protocol is especially good at large-scale signing because the signature rate does not drop dramatically like other CV QDS protocols when message size grows larger. Based on that, potential applications of multibit CV QDS can be conceived and outlooked.

The rest of this paper is organized as follows: Section \ref{sec2} provides an overview of our protocol, detailing the procedures of CV distribution method and the OTUH QDS scheme. In Sec. \ref{sec3}, we offer a comprehensive security proof of our protocol, analyzing both the CV procedure and the OTUH method applied in the QDS scheme. Section \ref{sec4} showcases numerical simulations conduced to evaluate the performance of our protocol and compare it with previous CV QDS schemes. Lastly, we draw conclusions and engage in a discussion regarding our work in Sect. \ref{sec5}.

\section{\label{sec2}Protocol Description}
In this section, we first introduce the schematic of our CV QDS protocol. In the simplest instance, we consider a signature scenario involving three parties: one sender, Alice, and two symmetric recipients, Bob and Charlie. Without loss of generality, we assume Bob to be the specified recipient, and Charlie automatically becomes the verifier. For a successful QDS scheme, Bob and Charlie should be able to determine that Alice is the genuine author of $m$, which means the forging and repudiation attacks are prevented. If all participants are honest, the scheme should succeed except with negligible probability.

As shown in Fig. \ref{fig1}, a basic communication network is constructed through quantum (solid lines) and classical (dashed lines) channels. Alice wishes to send a classical $m$-bit message to Bob, which he will forward to Charlie. This scheme could be easily extended to include more participants~\cite{arrazola2016multiparty}. We first provide a brief overview of the entire protocol.

\subsection{Protocol overview}
Our CV QDS protocol consists of two stages: a distribution stage and a messaging stage.

\begin{description}[leftmargin = 0 pt]
\item[Distribution stage]\
\begin{enumerate}[label = \alph*., leftmargin = 0 pt, itemindent = 25 pt]
    \item \textit{State preparation and measurement.} Alice generates random keys $\{X_a,Y_a\}$ and encodes them in coherent states, which are transmitted through quantum channels. Bob and Charlie measures the received states to get keys $\{X_b,Y_b\}$ and $\{X_b,Y_b\}$ respectively. The keys are asymmetric for the signer and receiver and satisfying perfect correlation conditions $X_a=X_b \oplus X_c$ and $Y_a=Y_b \oplus Y_c$.
    \item \textit{Parameter estimation.} Alice and Bob uses linear codes to reconcile their final keys. 
    \item \textit{Error correction.} Bob estimate some important parameters of the final keys like fidelity, bit error rate and phase error rate.
    \item \textit{Grouping.} Alice and Bob randomize the order of final keys and divide them into several $n$-bit strings for subsequent messaging task.
\end{enumerate}
\item[Messaging stage]\
\begin{enumerate}[label = \alph*., leftmargin = 0 pt, itemindent = 25 pt]
    \item \textit{Signing.} Alice generates the signature $Sig$ of a message $Doc$ with OTUH method using a key string generated in the distribution stage and transmits the packet $Pac=\{Doc, Sig\}$ containing the message and signature to Bob.
    \item \textit{Verification.} Bob forwards the packet to Charlie along with his keys $\{X_b,Y_b\}$, and Charlie sends his keys $\{X_c,Y_c\}$ in return on receipt. After the exchange, Bob and Charlie verify the integrity and authenticity of the message through a hashing test using the new calculated key strings $X_b \oplus X_c, Y_b \oplus Y_c\}$. The message and signature are accepted if and only if the OTUH tests success on both recipients' sides. 
\end{enumerate}
\end{description}

The subsequent contents of this section will elaborate on the steps of each stage in detail.

\subsection{Distribution stage}
In the distribution stage, Alice distributes coherent strings as shared keys to Bob and Charlie, respectively, by implementing CV method. To ensure security against general coherent attacks in finite-size regime, a cutting-edge fidelity test approach~\cite{chabaud2020building,matsuura2021finite} is employed during the parameter estimation process. In upcoming parts, we first introduce this approach and then outline complete steps of the distribution stage.

\begin{figure*}
    \centering
    \includegraphics[width=\linewidth]{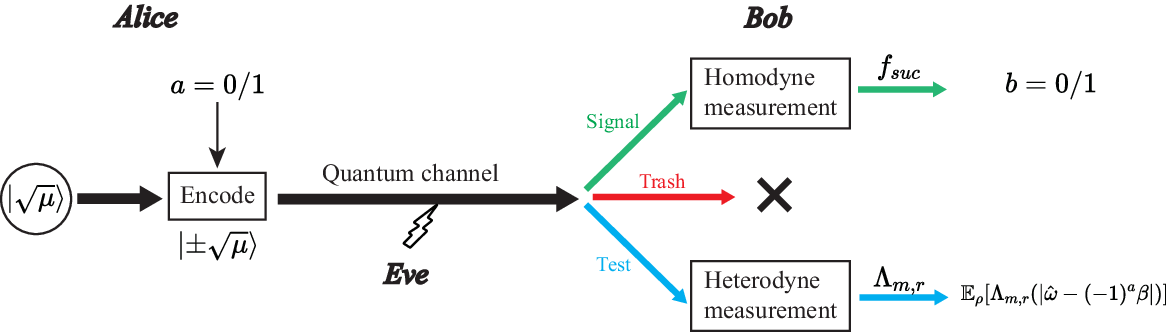}
    \caption{The schematic of key distribution based on CV. Alice generates a random bit $a \in \{0,1\}$ as her key and uses it to encode coherent states with amplitude $(-1)^a \sqrt{\mu}$. Then she transmit the states to Bob through a quantum channel with potential eavesdroppers Eve. Upon receiving, Bob use an optical switch to choose only one of three procedures based on the probabilities predetermined by practical conditions. In the signal round, Bob performs a homodyne measurement with an outcome $\hat{x}$ and use a acceptance function $f_{\text{suc}}$ to obtain his key $b \in \{0,1\}$. In the test round, he performs a heterodyne measurement with an outcome $\hat{\omega}$ and use the smooth function $\Lambda_{m,r}$ to calculate the bound of fidelity. In the trash round, he produces no outcome.}
    \label{fig2}
\end{figure*}

\subsubsection{Fidelity test approach}
The core element of our fidelity test approach is a smooth function that maps the unbounded outcome of heterodyne measurement to a bounded value. The smooth function is defined as follows.
\begin{defi}
	Let $\Lambda_{m,r}(\nu) (\nu \geqslant 0)$ be the smooth function given by
\begin{equation}
 	\Lambda_{m, r}(\nu):=e^{-r \nu}(1+r) L_{m}^{(1)}((1+r) \nu) 
\end{equation}
for an integer $m \geqslant 0$ and a real number $r > 0$. $L^{(k)}_n(\nu)$ represents associated Laguerre polynomials given by
\begin{equation}
 	L_{n}^{(k)}(\nu):=(-1)^{k} \frac{d^{k} L_{n+k}(\nu)}{d \nu^{k}},
\end{equation}
where
\begin{equation}
 	L_{n}(\nu):=\frac{e^{\nu}}{n !} \frac{d^{n}}{d \nu^{n}}\left(e^{-\nu} \nu^{n}\right) 
\end{equation}
are the Laguerre polynomials. The absolute value and the slope of the function are bounded with different parameters $(m,r)$. Through calculation in Ref.~\cite{matsuura2021finite}, we can adopt the optimized value $(m,r)=(1,0.4120)$ to minimize the range of the smooth function.
\end{defi}
Based on the definition, we can establish a lower bound on the fidelity of input pulses with a specified confidence level in finite-size regime. For fidelity to a coherent state $|\beta \rangle$, the inequality below holds.
\begin{equation}
	\label{e}
	\mathbb{E}_{\rho}\left[\Lambda_{m, r}\left(|\hat{\omega}-\beta|^{2}\right)\right] \leqslant \operatorname{Tr}(\rho |\beta\rangle \langle \beta|) \quad(m: odd).
\end{equation}
The proof of this inequality can be found in Appendix \ref{app_a}. Here we adopt the optimized value $(m,r)=(1,0.4120)$ to minimize the range of the smooth function. Upon that, we can obtain a measure of disturbance in the binary modulated CV scheme by monitoring the fidelity. Analogous to the bit errors in the B92 protocol~\cite{bennett1992quantum,tamaki2003unconditionally,koashi2004unconditional}, fidelity facilitates the construction of a security proof based on a reduction to entanglement distillation~\cite{shor2000simple,lo1999unconditional}, a technique commonly employed in DV schemes.

\subsubsection{Steps of key distribution}
Because of the symmetry of two recipients, Bob and Charlie, the key distribution procedures are identical between Alice and them respectively. For the simplicity of narration, we take Alice and Bob as example and illustrate the schematic of key distribution in Fig.\ref{fig2}. The detailed procedures are described below.
\begin{enumerate}[label = \arabic*., leftmargin = 0 pt, itemindent = 25 pt]
	\item \textit{Preparation.} Alice generates a random bit $a \in \{0,1\}$ and sends a coherent state with amplitude $(-1)^a\sqrt{\mu}$ to Bob. She repeats this process $N$ times. 
	\item \textit{Measurement.} For each received optical pulse, Bob chooses a label from $\{$signal, test, trash$\}$ with predetermined probabilities $p_{\text{sig}}, p_{\text{test}}$ and $p_{\text{trash}}$. According to the chosen label, Bob  use an optical switch to conduct only one of following procedures on the pulse respectively.
		\begin{description}[leftmargin = 0 pt]
			\item[signal] Bob performs a homodyne measurement on the received optical pulse, and obtains an outcome $\hat{x} \in \mathbb{R}$. With a predetermined probability function $f_{\text{suc}}(|\hat{x}|)$, which is ideally a step function, he regards the detection to be a ``success'' (otherwise ``failure''), and then announces this result of the detection. In the case of a success, he defines a bit b = 0 (1) when $sign(\hat{x})=+(-)$ and keeps $b$ as a sifted key bit which pairs with Alice's key bit $a$.
			\item[test] Bob performs a heterodyne measurement on the received optical pulse, and obtains an outcome $\hat{\omega}\in\mathbb{C}$. Alice announces her bit $a$ which makes Bob have the knowledge of the expected amplitude of his received state$(-1)^a\beta$. Then Bob calculates the value of $\Lambda_{m,r}(|\hat{\omega}-(-1)^a \beta|^2)$. 
			\item[trash] Alice and Bob produce no outcomes.
		\end{description}
		We refer to the numbers of ``success'' and ``failure'' signal rounds, test rounds, and trash rounds as $\hat{N}^{\text{suc}},\hat{N}^{\text{fail}}, \hat{N}^{\text{test}}$ and $\hat{N}^{\text{trash}}$, respectively ($N= \hat{N}^{\text{suc}}+\hat{N}^{\text{fail}}+\hat{N}^{\text{test}}+\hat{N}^{\text{trash}}$ holds by definition).
	\item \textit{Parameter estimation.} Bob calculates the sum of $\Lambda_{m,r}(|\hat{\omega}-(-1)^a \beta|^2)$ obtained in the $\hat{N}^{\text{test}}$ test rounds for the fidelity test, which is denoted by $\hat{F}$. For the bit error rate, Alice and Bob compare several bits to estimate.
	\item \textit{Error correction.} Alice and Bob consume part of keys through encrypted communication to perform following actions. Alice communicates Bob through linear codes for her sifted key and Bob reconciles his sifted key accordingly. We denote $H$ as the binary Shannon entropy, and then the consumption of this process can be estimated by $fH(E^b)\hat{N}^{\text{suc}}$ according to information theory, where $f$ is the error correction efficiency and $E^b$ represents the bit error rate of raw keys. Alice and Bob then verify the correction by comparing $\log_{2}{(1/{\epsilon_{\text{cor}}})}$ bits via universal hashing. Here $\epsilon_{\text{cor}}$ is the failure probability of this error correction process. 
	\item \textit{Grouping.} For the distilled $\hat{N}^{\text{fin}}$-bit keys, Alice then disturbs the orders of them randomly and publicizes the new order to Bob through the authenticated channel. Subsequently, Alice and Bob divide the final keys into several $n$-bit group, each of which is used for a signature task during the messaging stage. This procedure ensures that the attacker cannot predict the position of each bit within a specific group in advance.
\end{enumerate}
After all distribution process, the number of final keys is given by
\begin{equation}
    \hat{N}^{\text{fin}}=\hat{N}^{\text{suc}}[1-fH(E^b)]-\log_{2}{\frac{1}{\epsilon_{\text{cor}}}},
    \label{nfin}
\end{equation}
which consists of many $n$-bit groups.

It is worth noting that we omit the privacy amplification process, which is commonly employed in quantum communication. This omission would leads to final keys with full correctness and imperfect secrecy. Thanks to the specific utilization and requirements of distributed keys in QDS, recent research Ref.~\cite{li2023one} has proved that these imperfect keys are capable of realizing secure QDS protocols. Nevertheless, the leaked information available to Eve, i.e., the maximum unknown information of an $n$-bit group of keys, is still required when deciding group size $n$. The detailed formula will be deduced in the subsequent security analysis part.

To estimate the unknown information, the step of grouping makes a good entry point. In fact, the grouping process can be viewed as a form of random sampling in our finite-size analysis. Thus the unknown information of an $n$-bit group $\mathcal{H}_n$ after distribution stage can be bounded by
\begin{equation}
	\mathcal{H}_n \leqslant n[1-H(E^{\phi n})].
	\label{hn}
\end{equation}
Here, $E^{\phi n}$ is the phase error rate in each $n$-bit group which can be estimated based on the phase error rate of final keys $E^{\phi}$ and group size $n$. The detailed expression giving the upper bound of $E^{\phi n}$ can be found in Eq. (\ref{epn}) in Appendix \ref{app_b1}.
 
\subsection{Messaging stage}
In the messaging stage, we adopt the OTUH method which leverages the almost XOR universal (AXU) hash function to generate the signature for a long message. For the clarity of narration, we first introduce the AXU hash function and its properties.
\subsubsection{AXU hash function}
The AXU hash function is a special class of hash functions that map input values of arbitrary length to almost random hash values with a fixed length~\cite{carter1979universal}. The signature generated in OTUH QDS corresponds to the AXU hash value of the message to be signed, with the AXU hash function be determined by just one string of Alice keys. 

Various AXU hash can be employed in the messaging stage of our protocol depending on the diverse application scenarios of users. To demonstrate the messaging procedures in detail, we choose the linear feedback shift register-based (LFSR-based) Toeplitz hashing method~\cite{krawczyk1994lfsrbased} as our hash function.
\begin{defi}
	\label{lfsr}
	LFSR-based Toeplitz hash function: LFSR-based Toeplitz hash function can be expressed as $h_{p,s}(M) = H_{nm} \cdot M$ , where $p,s$ determines the function and $M$ is the message in the form of an $m$-bit vector. The detailed process of generating LFSR-based Toeplitz hash function is as follows: A randomly selected irreducible polynomial of order $n$ in the field GF(2), $p(x)$, determines the construction of LFSR. $p(x)=x^n + p_{n-1}x^{n-1} +\ldots +p_1 x+p_0$ can be characterized by its coefficients of order from zero to $n-1$, i.e., $p_a =(p^{n-1},p_{n-2},\ldots,p_1,p_0)$. For the initial state $s$ which is also represented as an $n$-bit vector $s=(a_n,a_{n-1},\ldots,a_2,a_1)^T$, the LFSR will be performed $m$ times to generate $m$ vectors. Specifically, it will shift down every element in the previous column, and add a new element to the top of the column. For example, the LFSR transforms $s$ into $s_1=(a_{n+1},a_n,\ldots,a_3,a_2)^T$, where $a_{n+1}=p_a \cdot s$, and likewise, transforms $s_1$ to $s_2$. Then the $m$ vectors $s,s_1,\ldots,s_{m-1}$ will together construct the Toeplitz matrix $H_{nm}=(s,s_1,\ldots,s_{m-1})$, and the hash value of the massage is $H_{nm} \cdot M$.
\end{defi}

\begin{figure}
    \centering
    \includegraphics[width=\linewidth]{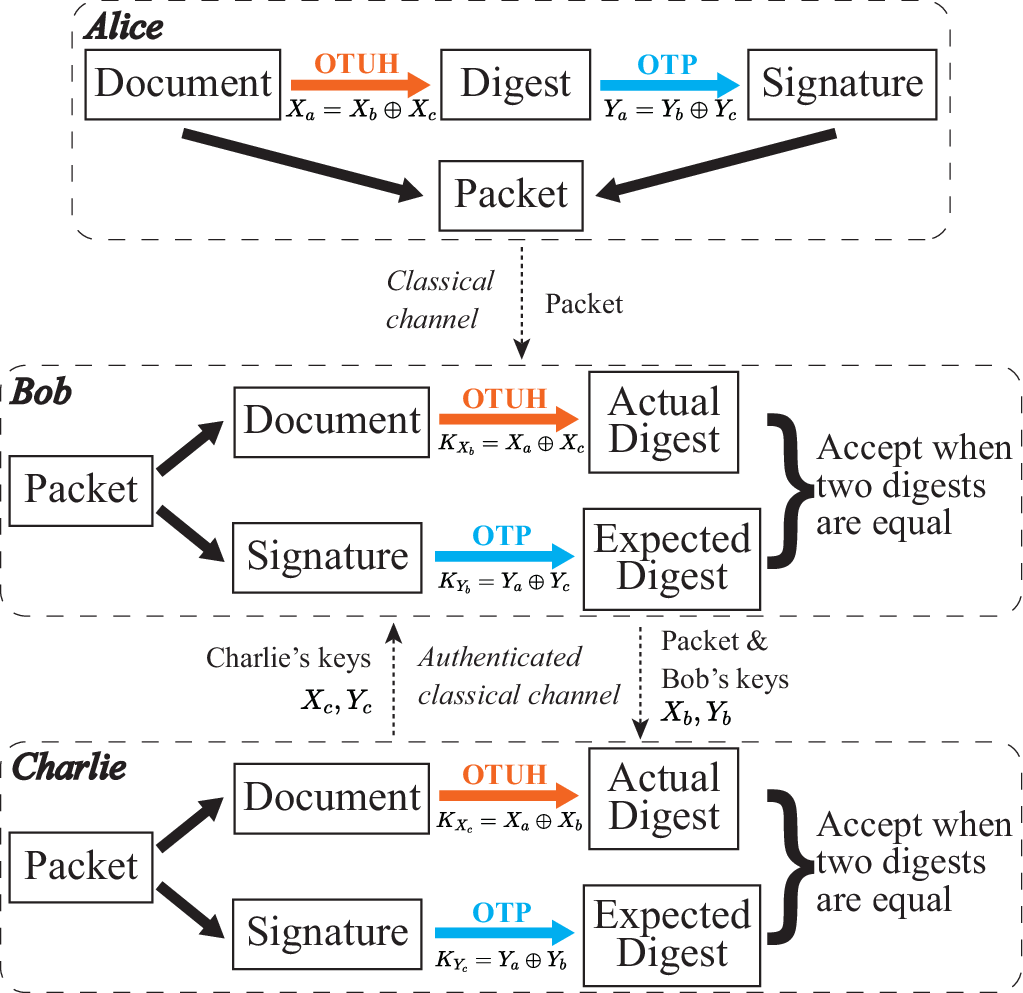}
    \caption{The schematic of OTUH QDS applied in the messaging stage. The OTUH (orange arrow) and OTP (blue arrow) method constitutes core steps in our QDS scheme to encode (Alice) and decode (Bob and Charlie) digest and signature, which guarantee the information-theoretical security of the protocol.}
    \label{fig3}
    \end{figure}

\subsubsection{Steps of messaging}
The schematic of OTUH QDS is shown is Fig. \ref{fig3}, we then describe steps adopting this method in messaging stage in detail.
\begin{enumerate}[label = \roman*., leftmargin = 0 pt, itemindent = 25 pt]
	\item \textit{Signing of Alice.} First, Alice uses a local quantum random number, characterized by an $n$-bit string $p_a$, to randomly generate an irreducible polynomial $p(x)$ of degree $n$~\cite{menezes1997handbook}. Second, she uses the initial vector (key bit string $X_a$) and irreducible polynomial (quantum random number $p_a$) to generate a random LFSR-based Toeplitz matrix $H_{nm}$, with $n$ rows and $m$ columns. Third, she uses a hash operation with $Hash= H_{nm} \cdot Doc$ to acquire an $n$-bit hash value of the $m$-bit document. Fourth, she exploits the hash value and the irreducible polynomial to constitute the $2n$-bit digest $Dig = (Hash||p_a)$. Fifth, she encrypts the digest with her key bit string $Y_a$ to obtain the $2n$-bit signature $Sig = Dig \oplus Y_a$ using OTP. Finally, she uses the public channel to send the message packet containing the signature and document $Pac=\{Sig, Doc\}$ to Bob.
	\item \textit{Verification of Bob.} Bob uses the authentication classical channel to transmit the received $Pac$, as well as his key bit strings $\{X_b, Y_b\}$, to Charlie. Then, Charlie uses the same authentication channel to forward his key bit strings $\{X_c, Y_c\}$ to Bob. Bob obtains two new key bit strings $\{K_{X_b} = X_b \oplus X_c, K_{Y_b} = Y_b \oplus Y_c\}$ by the XOR operation. Bob exploits $K_{Y_b}$ to obtain an expected digest and bit string $p_b$ via XOR decryption. Bob utilizes the initial vector $K_{X_b}$ and irreducible polynomial $p_b$ to establish an LFSR-based Toeplitz matrix. He uses a hash operation to acquire an n-bit hash value and then constitutes a $2n$-bit actual digest. Bob will accept the signature if the actual digest is equal to the expected. Then, he informs Charlie of the result. Otherwise, Bob rejects the signature and announces to abort the protocol.
	\item \textit{Verification of Charlie.} If Bob announces that he accepts the signature, Charlie then uses his original key and the key sent to Bob to create two new key bit strings $\{K_{X_c} = X_b \oplus X_c, K_{Y_c} = Y_b \oplus Y_c\}$. Charlie employs $K_{Y_c}$ to acquire an expected digest and bit string $p_c$ via XOR decryption. Charlie uses a hash operation to obtain an $n$-bit hash value and then constitutes a $2n$-bit actual digest, where the hash function is generated by initial vector $K_{X_c}$ and irreducible polynomial $p_c$. Charlie accepts the signature if the two digests are identical. Otherwise, Charlie rejects the signature.
\end{enumerate}
For the scheme with LFSR based Toeplitz hashing, a message of length $M$ requires Alice to generate six bit strings $X_b, X_c , Y_b, Y_c , Z_b, Z_c$, each of length $n$. Thus with the laser of repetition frequency $F$, we can write the signature rate formula as
\begin{equation}
    R=\frac{1}{3n} \frac{\hat{N}^{\text{fin}}}{F}.
    \label{rate}
\end{equation}
To achieve the best signature rate, we need to optimize parameters for large $\hat{N}^{\text{fin}}$ and small $n$, which constitute the core topic in our numerical simulation.

\section{\label{sec3}Security Analysis}
To establish the security proof of our protocol, two major aspects should be analyzed: the security of keys and the security of the QDS scheme. For the former, we embark on a distinct security analysis path, inspired by Ref.~\cite{matsuura2021finite,koashi2004unconditional}, which employs a qubit-based protocol to prove security against coherent attacks. During this analysis, we adopt mathematical techniques to meet the core requirement for estimating the phase error rate of the final keys. Regarding the QDS scheme, our analysis will encompass three fundamental security criteria, i.e., robustness, nonrepudiation and unforgeability in our analysis. These topics will be thoroughly discussed in their respective sections.

\begin{figure*}
    \centering
    \includegraphics[width=\linewidth]{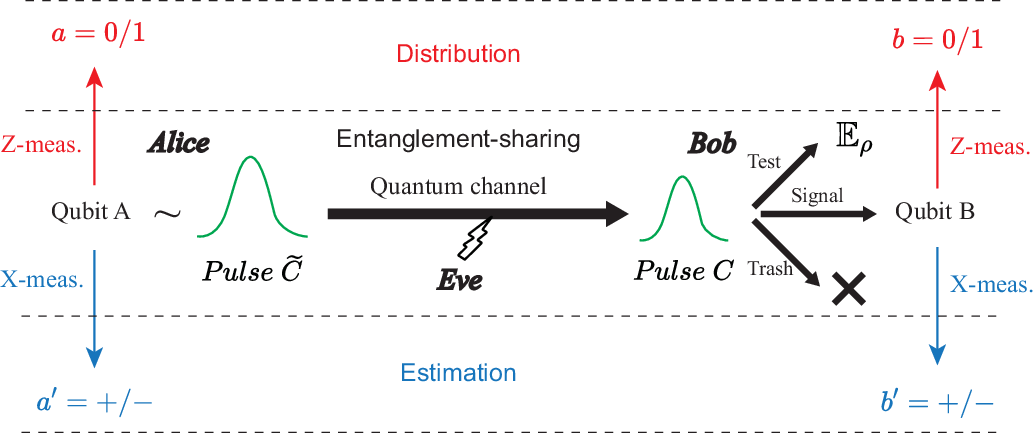}
    \caption{Relation between estimation and distribution procedures. The estimation (blue) and distribution (red) procedures are related through their common entanglement-sharing part. After the entanglement-sharing part, Alice and Bob are left with the observed data $(\hat{N}^{\text{suc}},\hat{N}^{\text{fail}}, \hat{N}^{\text{test}}, \hat{N}^{\text{trash}},\hat{F},\hat{Q}_{-})$ and $\hat{N}^{\text{suc}}$ pairs of qubits. If Alice and Bob ignore $\hat{Q}_{-}$ and measure their qubits on the $Z$-basis to determine their $\hat{N}^{\text{suc}}$-bit sifted keys, it becomes equivalent to the actual distribution procedure. On the other hand, if Alice and Bob measure their qubits on the $X$-basis, they can count the number of phase error $\hat{N}^{\text{suc}}_{\text{ph}}$, which we call the estimation procedure. If we can find an upper bound $U$ on $\hat{N}^{\text{suc}}_{\text{ph}}$, it restricts the property of $\hat{N}^{\text{suc}}$ pairs of qubits after entanglement-sharing, which in turn limits the leaked information on the sifted keys in the actual distribution procedure. Thus the security proof is reduced to finding a proper upper bound $U$ represented as a function of the variables that are commonly available in two procedures.}
    \label{fig4}
\end{figure*}

\subsection{Security of continuous-variable method}
In this section, we dissect our security analysis into two primary parts. Firstly, we convert the measurements of quadratures in our CV protocol into measurements of two-qubit entanglement in a qubit-based protocol, while the attacker Eve is unable to tell the difference. In other words, the qubit-based protocol is equivalent to our CV protocol in terms of security, which means we can utilize the existing conclusions of qubit-based protocol to prove the security against coherent attacks just like DV protocols~\cite{shor2000simple,hayashi2012concise}. As a link between these two protocols, we incorporate a coherent-based entanglement, a concept commonly employed in traditional CV analysis, to facilitate this transformation. After that, we carefully evaluate the phase error rate, a key parameter in the security analysis of qubit-based entanglement protocols, in the finite-size scenario. During the estimation process, we employ various mathematical techniques to achieve a suitable upper bound.
\subsubsection{Qubit entanglement based description}
The first task of our analysis is to establish a qubit entanglement model that accurately represents the preparation and measurement of optical pulses in our protocol. Through step 1 and 2 in the distribution stage, Alice and Bob share a pair of key bits $a$ and $b$ respectively for each success optical pulse. In traditional CV analysis, a coherent-based entanglement is employed to elucidate the relationship between key bits $a$ and $b$. This transformation originates from the widely acknowledged equivalence between the prepare-and-measure scheme and the entanglement-based scheme. To demonstrate this, we introduce a qubit $A$ that entangles with an optical pulse $\tilde{C}$ in a state
\begin{equation}
	\label{Psi}
 	|\Psi\rangle_{A \tilde{C}}:=\frac{|0\rangle_{A}|\sqrt{\mu}\rangle_{\tilde{C}}+|1\rangle_{A}|-\sqrt{\mu}\rangle_{\tilde{C}}}{\sqrt{2}}.
\end{equation}
With regard to this coherent-based entanglement, Alice retains qubit $A$ and Bob received the pulse as $C$. By conducting a measurement of the qubit $A$ on $Z$ basis $\{|0\rangle,|1\rangle\}$, Alice can determine the bit value $a$. Bob, on the other hand, can obtain the bit value $b$ through homodyne measurement on the pulse $C$. In this manner, we model the prepare-and-measure process using this coherent-based entanglement $|\Psi\rangle_{A \tilde{C}}$.

Utilizing this coherent-based entanglement as a bridge, we further construct a completely positive (CP) map for Bob defined by
\begin{equation}
 	\mathcal{F}_{C \rightarrow B}\left(\rho_{C}\right):=\int_{0}^{\infty} d x K^{(x)} \rho_{C} K^{(x) \dagger} 
\end{equation}
with
\begin{equation}
	\label{kx}
 	K^{(x)}:=\sqrt{f_{\text {suc}}(x)}(|0\rangle_{B}\langle x|_{C}+ |1\rangle_{B}\langle-x|_{C}),
\end{equation}
where $\langle x|$ maps a state vector to the value of its wave function at $x$. We assume that the pulse $C$ is in a state $\rho_C$ and the corresponding process succeeds with a probability $p_{\text{suc}}$, then the relation  $p_{\text{suc}}\rho_B=\mathcal{F}_{C \rightarrow B}(\rho_C)$ holds for the qubit $B$ in a state $\rho_B$. This means Bob probabilistically converts the received pulse $C$ to a qubit $B$. If the qubit $B$ is then measured on $Z$ basis, probabilities of the outcome $b = 0, 1$ are given by
\begin{equation}
 	p_{\text {suc}}\langle 0|\rho_{B}| 0\rangle=\int_{0}^{\infty} f_{\text {suc }}(x) d x\langle x|\rho_{C}| x\rangle,
\end{equation}
\begin{equation}
 	p_{\text {suc}}\langle 1|\rho_{B}| 1\rangle=\int_{0}^{\infty} f_{\text {suc}}(x) d x\langle-x|\rho_{C}|-x\rangle,
\end{equation}
which show that the $Z$-basis measurement outcomes of qubits $A$ and $B$ correspond to the sifted key bits $a$ and $b$. Thus far, we can say that the qubit entanglement between $A$ and $B$ is a complete representation of our CV protocol, which complete our qubit entanglement based description.

In fact, the entanglement based description outlined above is a mapping between the preparation and measurement of coherent states in a CV system and the shared entangled qubits in a DV system. The rationale behind this mapping is to adopt analysis techniques from DV to prove the security of our protocol. According to Preskill and Shor, a DV protocol is secure against coherent attacks with a sufficient amount of privacy amplification against leaked information, which is closely tied to the phase error rate of final keys. However, a direct measurement of the phase error rate is not feasible in our protocol, which pose a big challenge for us. To resolve this problem, our subsequent analysis will primarily concentrate on the estimation of the phase error rate.

To begin with, we define the phase error related to the qubit pair. Suppose that, after the preparation of qubits $A$ and $B$ previously discussed, Alice and Bob measure their $\hat{N}^{\text{suc}}$ pair of qubits on $X$ basis $\{|+\rangle,|-\rangle\}$ instead of $Z$ basis $\{|0\rangle,|1\rangle\}$. It is evident that occurrences of outcomes $(+,-)$ or $(-,+)$ represent phase errors. Then we can record the count of phase errors, denoted as $\hat{N}^{\text{suc}}_{\text{ph}}$, among the $\hat{N}^{\text{suc}}$ pairs. With this definition, our target is to acquire a reliable upper bound on the phase error rate $\hat{N}^{\text{suc}}_{\text{ph}}/\hat{N}^{\text{suc}}$, whose the binary Shannon entropy is sufficient for the estimation of the leaked information in the asymptotic limit.

To effectively address the finite-size scenario as well, a more stringent analysis of the upper bound for the phase error rate is required. Consequently, we introduce an estimation protocol that employs identical qubit entanglement distribution but distinct measurement configurations, as shown in Fig. \ref{fig4}. This approach leverages our previous findings to showcase the effectiveness of our estimation techniques for the phase error rate.
\begin{description}[leftmargin = 0 pt]
\item[Estimation protocol]\ 
\begin{enumerate}[label = \arabic*'., leftmargin = 0 pt, itemindent = 25 pt]
	\item \textit{Preparation.} Alice prepares a qubit $A$ and an optical pulse $\tilde{C}$ in an entangled state $|\Psi \rangle_{A\widetilde{C}}$ defined in Eq. (\ref{Psi}). She repeats it $N$ times.
	\item \textit{Measurement.} For each received pulse, Bob announces a label in the same way as that in Step 2. According to the chosen label, Alice and Bob do one of the following procedures. 
		\begin{description}[leftmargin = 0 pt]
			\item[signal] Bob performs a quantum operation on the received pulse $C$ specified by the CP map $\mathcal{F}_{C\rightarrow B}$ to determine success or failure of the detection and obtain a qubit $B$ upon success. He announces success or failure of detection. In the case of a success, Alice keeps her qubit $A$.
			\item[test] Bob performs a heterodyne measurement on the received optical pulse $C$, and obtains an outcome $\hat{\omega}$. Alice measures her qubit $A$ on $Z$ basis and announces the outcome $a \in \{0, 1\}$. Bob calculates the value of $\Lambda_{m,r}(|\hat{\omega}-(-1)^a \beta|^2)$ as defined in Step 2.
			\item[trash] Alice measures her qubit $A$ on $X$ basis to obtain $a_t \in \{+,-\}$.
		\end{description}
	Here $\hat{N}^{\text{suc}},\hat{N}^{\text{fail}}, \hat{N}^{\text{test}}, \hat{N}^{\text{trash}}$ and $\hat{F}$ are defined in the same way as those in Step 2. Let $\hat{Q}_{-}$ be the number of rounds in the $\hat{N}^{\text{trash}}$ trash rounds with $a_t=-$.
	\item \textit{Parameter Estimation.} Alice and Bob measure each of their $\hat{N}^{\text{suc}}$ pairs of qubits on $X$ basis and obtain outcomes $a'$ and $b'$, respectively. Let $\hat{N}^{\text{suc}}_{\text{ph}}$ be the number of pairs found in the combination $(a',b')=(+,-)$ or $(-,+)$.
\end{enumerate}
\end{description}
After such definition, we actually obtain pairs of key bits $a'$ and $b'$ that contains the information of phase error about the originally shared qubit entanglement $A$ and $B$. Here it's beneficial for us to make some clarification about what property Bob could get through his $X$-basis measurement in this estimation protocol. Let $\Pi_{\text{ev(od)}}$ be a projection operator to the subspace with even (odd) photon numbers. ($\Pi_{\text{ev}} + \Pi_{\text{od}} = \mathbb{I}_C$ holds by definition.) There is simple property that $\Pi_{\text{ev}} - \Pi_{\text{od}}$ is the operator for an optical phase shift of $\pi$, so we have $(\Pi_{\text{ev}} - \Pi_{\text{od}})|x\rangle = |-x\rangle$. Then Eq. (\ref{kx}) can be rewritten as
\begin{equation}
 	K^{(x)}=\sqrt{2 f_{\mathrm{suc}}(x)} (|+\rangle_{B}\langle x|_{C} \Pi_{\mathrm{ev}}+\mid- \rangle_{B} \langle x |_{C} \Pi_{\mathrm{od}}). 
\end{equation}
Therefore, when the state of the pulse $C$ is $\rho_C$, the probability of obtaining $+(-)$ in the $X$-basis measurement in the estimation protocol is given by
\begin{equation}
 	\left\langle+(-)\left|\mathcal{F}_{C \rightarrow B}\left(\rho_{C}\right)\right|+(-)\right\rangle=\operatorname{Tr}\left(\rho_{C} M_{\mathrm{ev}(\mathrm{od})}^{\mathrm{suc}}\right),
\end{equation}
where
\begin{equation}
 	M_{\mathrm{ev}(\mathrm{od})}^{\mathrm{suc}}:=\int_{0}^{\infty} 2 f_{\mathrm{suc}}(x) d x \Pi_{\mathrm{ev}(\mathrm{od})}|x\rangle \langle x|_{C} \Pi_{\mathrm{ev}(\mathrm{od})}.
\end{equation}
This equation means that the sign of Bob’s $X$-basis measurement outcome distinguishes the parity of the photon number of the received pulse. In this sense, the secrecy of final keys is assured by the complementarity between these two characteristics.

\subsubsection{Upper bound of phase error rate}
Before delving into detailed estimation techniques, it is imperative to outline the security conditions associated with phase errors. In recent researches~\cite{koashi2009simple,hayashi2012concise,matsuura2019refined}, the security proving task of the actual protocol is effectively reduced to construction of a function $U(\hat{F},\hat{N}^{\text{trash}})$ which satisfies
\begin{equation}
	\label{pr}
 	\operatorname{Pr}\left[\hat{N}_{\mathrm{ph}}^{\mathrm{suc}} \leqslant U\left(\hat{F}, \hat{N}^{\text {trash }}\right)\right] \geqslant 1-\epsilon
\end{equation}
for any attack towards the estimation protocol in finite-size regime. In essence, the inequality stipulated above constitutes the security condition for the final keys during the actual distribution phase. For the sake of conciseness, we will not elaborate further on this topic. The comprehensive definitions and rigorous proofs of security can be found in the references cited.

Learned from condition (\ref{pr}), function $U(\hat{F},\hat{N}^{\text{trash}})$ can obviously be constructed as the upper bound for phase errors. As an intermediate step toward our goal, we can first derive a bound on the expectation value $\mathbb{E}[\hat{N}^{\text{suc}}_{\text{ph}}]$ in terms of those collected in the test and the trash rounds, $\mathbb{E}[\hat{F}]$ and $\mathbb{E}[\hat{Q}_{-}]$, in the estimation protocol. We define relevant operators as
\begin{equation}
 	M_{\mathrm{ph}}^{\mathrm{suc}}:=|+\rangle\langle+|_{A} \otimes M_{\mathrm{od}}^{\mathrm{suc}}+\mid-\rangle\langle-|_{A} \otimes M_{\mathrm{ev}}^{\mathrm{suc}},
\end{equation}
\begin{equation}
 	\Pi^{\mathrm{fid}}:=|0\rangle\langle 0|_{A} \otimes\beta\rangle\langle\beta|_{C}+ 1\rangle\langle 1|_{A} \otimes -\beta\rangle\langle-\beta|_{C},
\end{equation}
\begin{equation}
 	\Pi_{-}^{\text {trash }}:=|-\rangle\langle-|_{A} \otimes \mathbb{I}_{C}.
\end{equation}
Then we immediately have
\begin{equation}    \mathbb{E}[\hat{N}_{\mathrm{ph}}^{\mathrm{suc}}]=p_{\mathrm{sig}} N \operatorname{Tr}(\rho_{A C} M_{\mathrm{ph}}^{\mathrm{suc}}),
\end{equation}
\begin{equation}
 	\mathbb{E}[\hat{F}] \leqslant p_{\text {test }} N \operatorname{Tr}(\rho_{A C} \Pi^{\text {fid}}),
\end{equation}
\begin{equation}
 	\mathbb{E}[\hat{Q}_{-}]=p_{\text {trash}} N \operatorname{Tr}(\rho_{A C} \Pi_{-}^{\text {trash }}).
\end{equation}
Here we apply Eq. (\ref{e}) to derive the inequality above. For simplicity, we denote $\operatorname{Tr}(\rho_{AC}M)$ as $\langle M \rangle$ for any operator $M$. The set of points $(\langle M^{\text{suc}}_{\text{ph}} \rangle,\langle \Pi^{\text{fid}} \rangle, \langle \Pi^{\text{trash}}_{-} \rangle)$ for all the density operators $\rho_{AC}$ form a convex region. Rather than directly deriving the boundary of the region, it is easier to pursue linear constraints in the form of
\begin{equation}
    \label{M}
 	\left\langle M_{\mathrm{ph}}^{\mathrm{suc}}\right\rangle \leqslant B(\kappa, \gamma)-\kappa\left\langle\Pi^{\mathrm{fid}}\right\rangle+\gamma\left\langle\Pi_{-}^{\mathrm{trash}}\right\rangle,
\end{equation}
where $B(\kappa,\gamma),\kappa,\gamma \in \mathbb{R}$. It is expected that a meaningful bound is obtained only for $\kappa, \gamma \geqslant 0$, because decreasing fidelity $\left\langle\Pi^{\mathrm{fid}}\right\rangle$ and stronger pulse, which directly increases $\left\langle\Pi_{-}^{\mathrm{trash}}\right\rangle$, would lead to a larger value of phase error rate. 

To find a function $B(\kappa, \gamma)$ satisfying Eq. (\ref{M}), let us define an operator
\begin{equation}
 	M[\kappa, \gamma]:=M_{\mathrm{ph}}^{\mathrm{suc}}+\kappa \Pi^{\mathrm{fid}}-\gamma \Pi_{-}^{\mathrm{trash}}.
\end{equation}
Then Eq. (\ref{M}) is rewritten as $\operatorname{Tr}(\rho_{AC}M[\kappa,\gamma]) \leqslant B(\kappa,\gamma)$. This condition should hold for all $\rho_{AC}$ if and only if $M[\kappa, \gamma]$ satisfies an operator inequality
\begin{equation}
	\label{ineq}
 	M[\kappa,\gamma] \leqslant B(\kappa, \gamma) \mathbb{I}_{AC}. 
\end{equation}
Theoretically speaking, although $B(\kappa,\gamma)=\sigma_{\text{sup}}(M[\kappa,\gamma])$ would give the tightest bound for the operator inequality above, where $\sigma_{\text{sup}}(O)$ denotes the supremum of the spectrum of a bounded self-adjoint operator $O$(i.e., the maximum modulus of eigenvalues of a matrix), it is difficult to compute it numerically since $M[\kappa, \gamma]$ has an infinite rank. As a compromise for the computable but not necessarily tight bound, we reduce the problem to replacing $M[\kappa, \gamma]$ with a constant upper bound except in a relevant finite-dimensional subspace spanned by $|\pm \beta \rangle$ and $M^{\text{suc}}_{\text{ev(od)}}|\pm \beta \rangle$. Then we just need to calculate the eigenvalues of some small-size matrices to get $B(\kappa,\gamma)$. For the detailed expression of $B(\kappa,\gamma)$, see Appendix \ref{app_b2}.

With $B(\kappa, \gamma)$ computed, we can rewrite Eq. (\ref{M}) to obtain
\begin{equation}
	\label{T}
 	\mathbb{E}[\hat{T}[\kappa, \gamma]] \leqslant N B(\kappa, \gamma),
\end{equation}
where $ \hat{T}[\kappa, \gamma]:=p_{\text {sig }}^{-1} \hat{N}_{\mathrm{ph}}^{\mathrm{suc}}+p_{\text {test }}^{-1} \kappa \hat{F}-p_{\text {trash }}^{-1} \gamma \hat{Q}_{-} $. This relation directly leads to an explicit bound on the phase error rate as
\begin{equation}
 	\mathbb{E}\left[\hat{N}_{\mathrm{ph}}^{\mathrm{suc}}\right] / p_{\mathrm{sig}} N \leqslant B(\kappa, \gamma)+\gamma q_{-}-\kappa \mathbb{E}[\hat{F}] / p_{\text {test }} N,
\end{equation}
which is enough for the computation of asymptotic key rates. Here $q_-$ is short for $\mathbb{E}[\hat{Q}_-]$.

As for the finite-size regime, the proof could be provided as follows. We use Azuma’s inequality~\cite{azuma1967weighted} to evaluate the fluctuations around the expectation value, leading to an inequality revised from Eq. (\ref{T})
\begin{equation}
	\label{delta1}
 	\hat{T}[\kappa, \gamma] \leqslant N B(\kappa, \gamma)+\delta_{1}(\epsilon/2),
\end{equation}
which holds with a probability no smaller than $1-\epsilon/2$ (see Appendix \ref{app_b3} for detailed expression of function $\delta_1$).

Another revision of proof happens in the definition of $\hat{T}[\kappa, \gamma]$. It includes $\hat{Q}_{-}$ which is inaccessible in the actual distribution stage, but we can derive bound by noticing that it is an outcome from Alice’s qubits and independent of the adversary’s attack. In fact, given $\hat{N}_{\text{trash}}$, it is the tally of $\hat{N}_{\text{trash}}$ Bernoulli trials with a probability $q_{-}$. Hence, we can derive an inequality of the form
\begin{equation}
	\label{delta2}
 	\hat{Q}_{-} \leqslant q_{-} \hat{N}^{\text {trash }}+\delta_{2}\left(\epsilon / 2 ; \hat{N}^{\text {trash }}\right),
\end{equation}
which holds with a probability no smaller than $1 - \epsilon/2$. Here $\delta_{2}(\epsilon/2; \hat{N}^{\text{trash}})$ can be determined by a Chernoff bound (see Appendix \ref{app_b3} for detailed expression of function $\delta_2$). Combining Eqs. (\ref{delta1}) and (\ref{delta2}), we obtain $U(\hat{F}, \hat{N}^{\text{trash}})$ satisfying Eq. (\ref{pr}) as 
\begin{equation}
 	\begin{aligned} 
 		U(\hat{F}, \hat{N}^{\text {trash }})= & p_{\text {sig }} N B(\kappa, \gamma)+p_{\text {sig }} \delta_{1}(\epsilon / 2) -\frac{p_{\text {sig }}}{p_{\text {test }}} \kappa \hat{F} \\
 		& +\frac{p_{\text {sig }}}{p_{\text {trash }}} \gamma(q_{-} \hat{N}^{\text {trash }}+\delta_{2}(\epsilon / 2 ; \hat{N}^{\text {trash }})).
 	\end{aligned} 
\end{equation}
With this construction, condition (\ref{pr}) is properly satisfied, which completes our security proof in finite-size regime. To conclude, we further give the upper bound of phase error rate as 
\begin{equation}
	E^{\phi}\leqslant U(\hat{F}, \hat{N}^{\text {trash }})/ p_{\mathrm{sig}} N.
	\label{ep}
\end{equation}

\subsection{Security of QDS}
In this part, we first focus on the properties of AXU hashing in the context of imperfect keys with limited secrecy leakage and then discuss the security of the entire QDS system.

\subsubsection{Attacks on AXU hashing}
There are various attacks available for the attacker to perform towards the AXU hashing method. The worst strategy is to randomly generate $m,t$ to satisfy $h_{p,s}(m)=t$, whose success probability is only $2^{-n}$. We leave out this attack in the following analysis. Apart from this, remaining attacks includes guessing keys and recovering keys from signatures. For the latter strategy, as the signature is encrypted by the key strings, the attacker must guess the key strings before the recovering algorithm. This will reduce its success probability to no more than that of just guessing keys. Therefore, the optimal strategy on AXU hashing is to directly guess the key string that encrypts the polynomial. In the following paragraphs, we will analyze the success probability of this strategy along the lines provided by Ref.~\cite{li2023one}.
 
 First, we quantify the guessing probability of the attacker when he is to guess an $n$-bit key string with limited secrecy leakage. Suppose the $n$-bit key string is $X$ and the attacker’s system is $E$, general attacks are modeled as a process where attackers can perform any operations on the system of all quantum states, get a system $\rho_E^x$ and perform any positive operator-valued measure $\{M_E^x\}_x$ performed on it. The probability that the attacker correctly guesses $X$ when in optimal situation is denoted as $P_{\text{guess}}(X|E)$. According to the definition of min-entropy~\cite{konig2009operational}, it naturally leads to
\begin{equation}
    P_{\text {guess }}(X|E)=\max _{\left\{M_{E}^{x}\right\}_{x}} \sum_{x} P_{x} \operatorname{Tr}\left(M_{E}^{x} \rho_{E}^{x}\right)=2^{-H_{\min }(X \mid E)_{\rho}}, 
    \label{pg}
\end{equation}
where $H_{\min}(X|E)_{\rho}$ is the min-entropy of $X$ and $E$. If $X$ is generated in the distribution stage of our protocol, the min-entropy can be further estimated by $H_{\min}(X|E)_{\rho}=\mathcal{H}_n$.

Next, we provide detailed analysis about the LFSR-based Toeplitz hash function (see Definition \ref{lfsr}) used in our protocol. Despite the utilization of two different keys, $X_a$ and $Y_a$, in the hash function, we could show that only guessing $X_a$ is enough to perform an effective attack. 

Suppose the attacker obtains a string $X_g$ as his guess of $X_a$. The optimal success possibility of this can be given by Eq. (\ref{pg}) as
\begin{equation}
    P(X_{a}=X_{g})=2^{-\mathcal{H}_{n}}.
\end{equation}
Thereafter, he decrypt it to obtain $p_g$ as his guessing of $p_a$ and transform $p_g$ into a polynomial $p_g(x)$ of order $n$. If a $m$-bit string $g$ and its matching polynomial $g(x)$ is generated to satisfy $p_g(x)|g(x)$, it can be proved (see Appendix. \ref{app_c}) that there is the relationship $h(g) = 0$ if $p_g = p_a$ (or equivalently $X_g = X_a$). This construction is simple since the attacker can select no more than $m/n$ polynomials and multiply them to constitute his choice of $g(x)$ of order $m$, which means he would guess the string $X_a$ for no more than $m/n$ times. Additionally, it must be noted that the attacker knows $p_a$ is irreducible, so he will only choose those guesses that satisfy this condition. The success probability of this optimized strategy can be expressed as
\begin{equation}
    P_{1}=\frac{m}{n} P\left(X_{a}=X_{g} \mid p_{g} \in \mathcal{I}\right), 
\end{equation}
where $P(A|B)$ represents conditional probability and $\mathcal{I}$ is the set of all irreducible polynomials of order $n$ in GF(2). The cardinal number of $\mathcal{I}$, i.e., the number of all irreducible polynomials of order $n$ in GF(2), is more than $2^{n-1}/n$. Thus $P(p_g \in \mathcal{I}) \leqslant (2^{n-1}/n)/2^n = 1/2n$. It is obvious that $P (X_a = X_g, p_g \in \mathcal{I}) = P(X_a = X_g)$ because $X_a = X_g$ assures $p_g = p_a \in \mathcal{I}$. Finally, we can get the estimation of the success probability of this kind of attack as
\begin{equation}
 	\begin{aligned}
 	 	P_{1} & =\frac{m}{n} \frac{P\left(X_{a}=X_{g}\right)}{P\left(p_{g} \in \mathcal{I}\right)} \\ 
 	 	& \leqslant \frac{m}{n} \frac{2^{-\mathcal{H}_{n}}}{\frac{1}{2 n}} =2^{1-\mathcal{H}_{n}} m \equiv \epsilon_{\mathrm{LFSR}}.
 	 	\label{lsfr}
 	 \end{aligned} 
\end{equation}
Here we denote the quantified upper bound above as $\epsilon_{\mathrm{LFSR}}$. In other words, it bounds the failure probability of authentication based on LFSR-based Toeplitz hashing with imperfect keys. According to the security condition, this value should be less than the security bound $\epsilon_s$, which gives
\begin{equation}
    2^{1-\mathcal{H}_n} m \leqslant \epsilon_s.
	\label{nbound}
\end{equation}
Combined with Eq. (\ref{hn}), we can obtain the inequality
\begin{equation}
	1-\log_{2} (\epsilon_s/m)\leqslant n[1-H(E^{\phi n})],
\end{equation}
from which the smallest group size $n$ can be thus calculated.

\subsubsection{Security of QDS scheme}
In a QDS scheme, the security analysis contains three parts: robustness, non-repudiation and unforgeability.
\begin{figure*}
    \begin{minipage}[t]{0.49\textwidth}
        \centering
        \includegraphics[width=\linewidth]{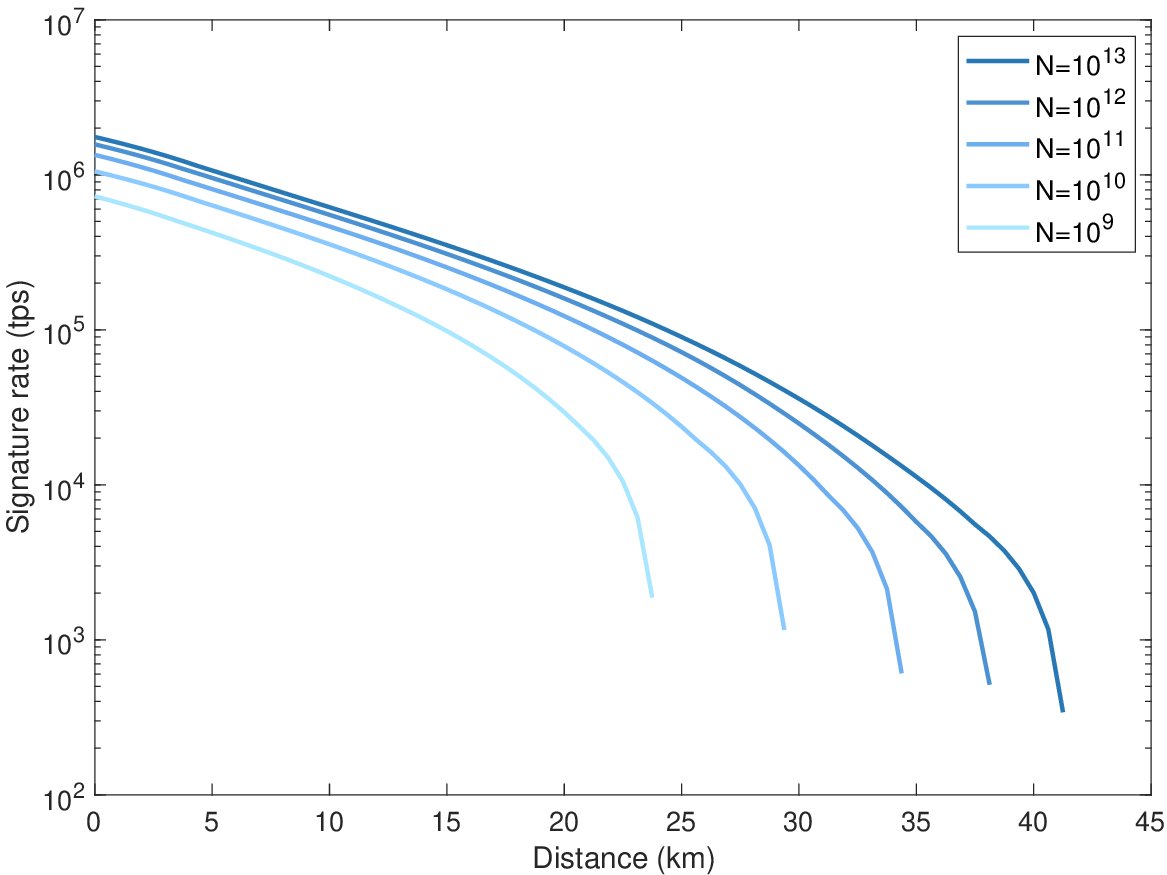}
        \caption{Signature rates of our protocol under different data sizes $10^{13}, 10^{12}, 10^{11}, 10^{10}$ and $10^{9}$. The amount of excess noise is set to zero. The message size is assumed to be 1 kb and the repetition rate of the laser is 1 GHz. The security bound is $10^{-10}$}
        \label{compare1}
    \end{minipage}  
    \hfill
    \begin{minipage}[t]{0.49\textwidth}
        \centering
        \includegraphics[width=\linewidth]{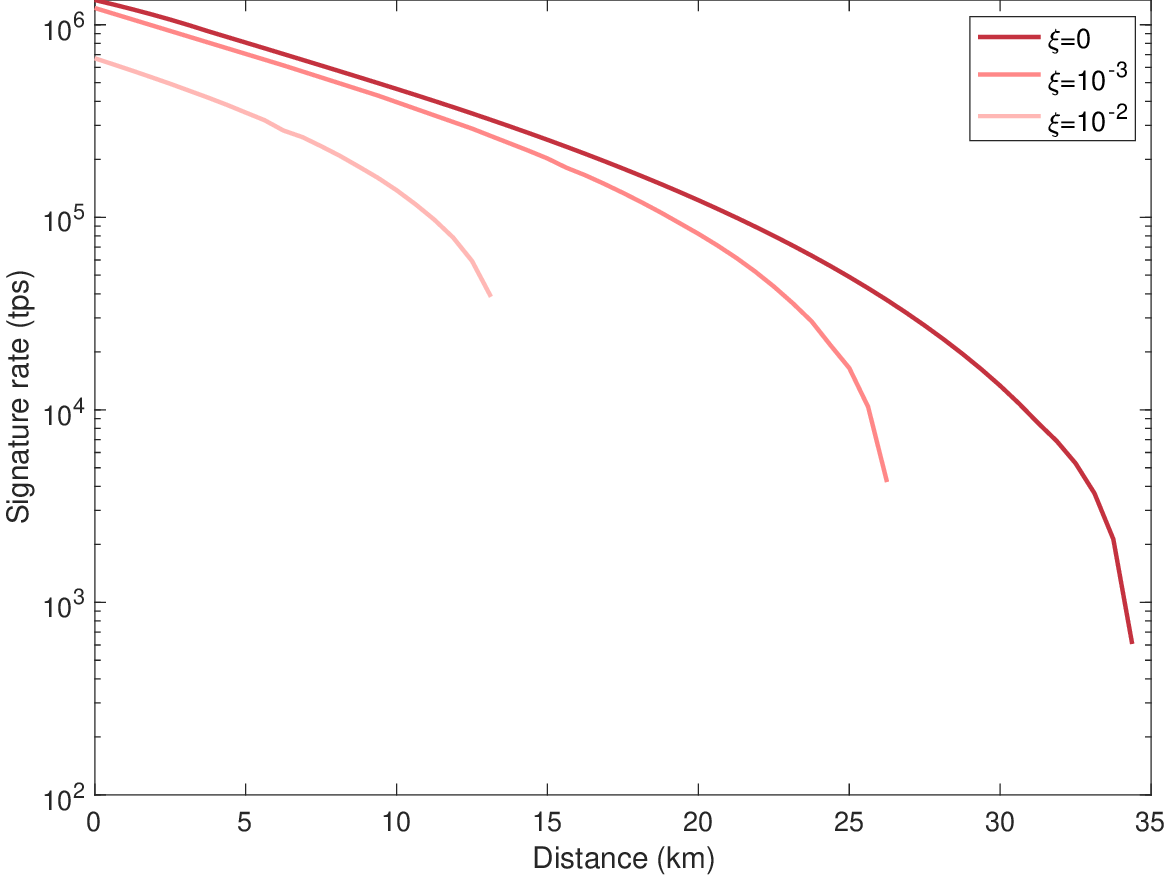}
        \caption{Signature rates of our protocol with different amount of excess noise $\xi = 0, 10^{-3} $ and $10^{-2}$. The data size is $10^{11}$. The message size is assumed to be 1 kb and the repetition rate of the laser is 1 GHz. The security bound is $10^{-10}$}
        \label{compare2}
    \end{minipage}
\end{figure*}
\begin{enumerate}[label = (\roman*), leftmargin = 0 pt, itemindent = 25 pt]
	\item \textit{Robustness.} The abortion in an honest run occurs only when Alice and Bob (or Charlie) share different key strings after the distribution stage. In our protocol, Alice and Bob (Charlie) will perform error correction in the distribution stage, which makes them share the identical final key. Thus, the robustness bound is $\epsilon_{\mathrm{rob}}=2\epsilon_{\mathrm{cor}}+2\epsilon'$, where $\epsilon_{\mathrm{cor}}$ is the failure probability of error correction and $\epsilon'$ is the probability that error occurs in classical message transmission. Here we assume $\epsilon' = 10^{-11}$ for simplicity.
	\item \textit{Nonrepudiation.} Alice successfully repudiates when Bob accepts the message while Charlie rejects it. For Alice’s repudiation attacks, Bob and Charlie should be both honest. From the perspective of Alice, Bob and Charlie’s keys are totally symmetric and would lead to the same decision for the same message and signature. Thus, Alice’s repudiation attack succeed only when error occurs in one of the key exchange steps, i.e., the repudiation bound is $\epsilon_{\mathrm{rep}}=2\epsilon'$.
	\item \textit{Unforgeability.} Bob forges successfully when Charlie accepts the forged message forwarded by Bob. After distribution stage, Bob (Charlie) can obtain no information of Alice’s keys which decides the AXU hash function before he transfer the message and signature to Charlie (Bob). Then Bob’s (Charlie's) forging attack in this protocol is equivalent to the attack attempting to forge the authenticated information sent from Alice to Charlie. Therefore, the probability of a successful forgery can be determined by the failure probability of hashing, i.e., one chooses two distinct messages with identical hash values. For the scheme utilizing LFSR-based Toeplitz hash , the bound is $\epsilon_{\mathrm{for}}=2^{1-\mathcal{H}_n}m$.
\end{enumerate}
In conclusion, the total security bound of QDS, i.e., the maximum failure probability of the protocol, is $\epsilon_{\mathrm{QDS}}=\max\{\epsilon_{\mathrm{rob}},\epsilon_{\mathrm{rep}},\epsilon_{\mathrm{for}}\}$.

\section{\label{sec4}Results}
To summarize, our protocol distributes $\hat{N}^{\text{fin}}$ keys in the distribution stage and estimate the unknown information $\mathcal{H}_n$. In the grouping process, the group size $n$ is decided by the security bound condition $\epsilon_{\text{LFSR}} \leqslant \epsilon_{s}$. During the messaging signing, the signature rate can be calculated by Eq. (\ref{rate}).

As a demonstration of the performance of our protocol, we conduct a simulation to calculate the signature rate with different parameters. In subsequent contents, we first introduce the simulation model for our numerical calculation and then make comparisons with different conditions and other protocols. 

\begin{figure*}
    \begin{minipage}[t]{0.49\textwidth}
        \centering
        \includegraphics[width=\linewidth]{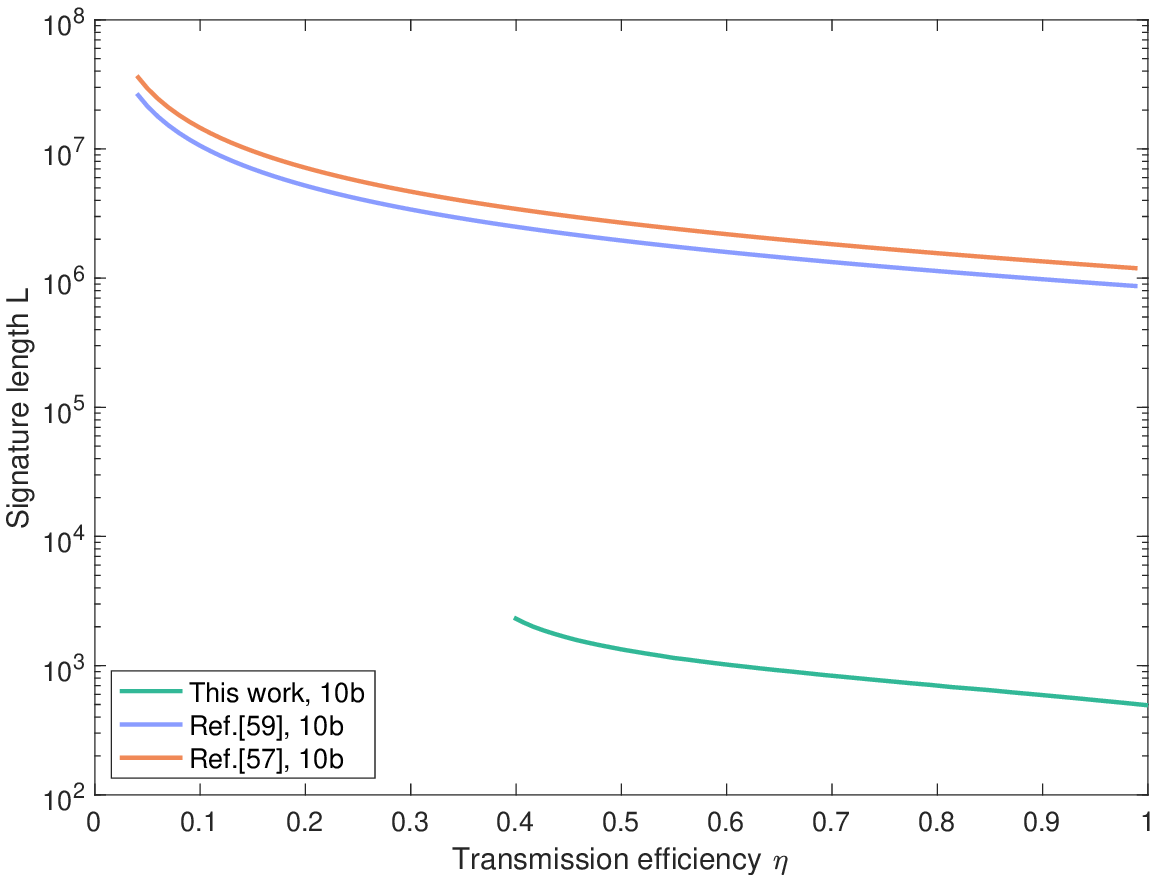}
        \caption{Signature length $L$ required to sign 10b message in our protocol, Ref.~\cite{zhang2019high} and Ref.~\cite{zhao2021multibit} against transmission efficiency $\eta$. The distances between Alice-Bob and Alice-Charlie are assumed to be the same. The data size N is $10^{11}$ and excess noise is $\xi=10^{-3}$. The security bound is $10^{-10}$.}
        \label{compare3}
    \end{minipage}  
    \hfill
    \begin{minipage}[t]{0.49\textwidth}
        \centering
        \includegraphics[width=\linewidth]{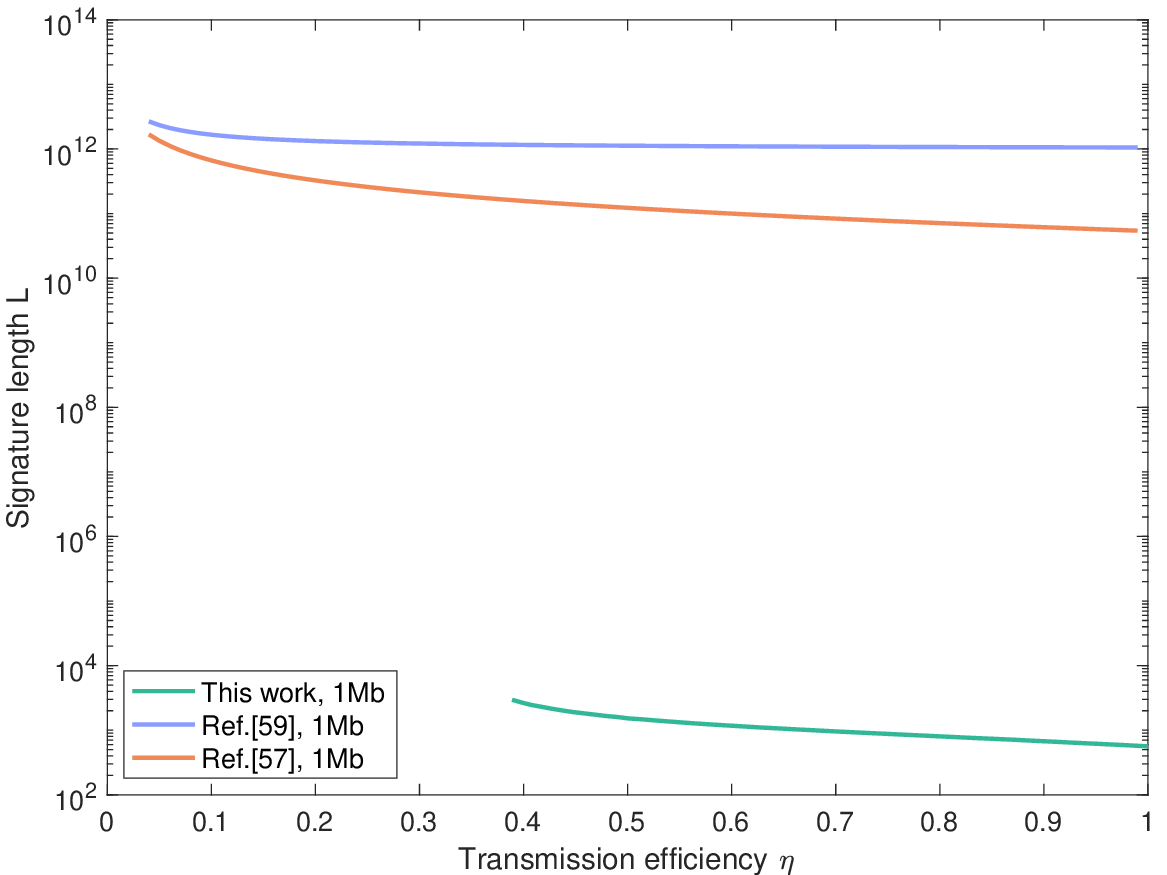}
        \caption{Signature length $L$ required to sign 1Mb message in our protocol, Ref.~\cite{zhang2019high} and Ref.~\cite{zhao2021multibit} against transmission efficiency $\eta$. The distances between Alice-Bob and Alice-Charlie are assumed to be the same. The data size N is $10^{11}$ and excess noise is $\xi=10^{-3}$. The security bound is $10^{-10}$.}
        \label{compare4}
    \end{minipage}
\end{figure*}

\subsection{Numerical simulation}
We assume a quantum channel model characterized by transmission efficiency $\eta$ and an excess noise $\xi$ at the channel output, which applies equally to both the Alice-Bob and Alice-Charlie channels with the same distance. Under this condition, the states Bob receives are obtained by randomly displacing coherent states $|\pm \eta \mu \rangle$ to increase their variances by a factor of $(1+\xi)$. Thus the expected amplitude of coherent state, denoted as $\beta$, is $\sqrt{\eta \mu}$. It is worth noting that these two channel parameters need to be estimated in practical system, a brief discussion of which will be provided in Sec. \ref{sec5}. In simulations, they will be treated as independent variables to demonstrate the performance of our protocol under various scenarios. For other protocol parameters, we set the repetition rate of the laser is 1 GHz and take a step function with a threshold $x_{\text{th}}( > 0)$ as the acceptance probability $f_{\text{suc}}(|x|)$ given in Eq. (\ref{fsuc}). For the fidelity test, we adopted $m = 1$ and $r = 0.4120$, which leads to $(\max \Lambda_{m,r},\min \Lambda_{m,r})=(2.824,-0.9932)$. Also, we take parameters $\epsilon_{\text{cor}}=2^{-51}$ and $\epsilon=2^{-104}$. 

With these preliminary settings, we are tasked with determining two auxiliary coefficients $(\kappa,\gamma)$ and five protocol parameters $(\mu,x_{\text{th}},p_{\text{sig}},p_{\text{test}},n)$. For each data size $N$, transmission efficiency $\eta$ and excess noise $\xi$ value, we could estimate bit error rate $E^b$ and phase error rate $E^\phi$ of distributed keys through Eqs. (\ref{eb}) and (\ref{ep}). Utilizing these estimates, we can then calculate the number of final keys $\hat{N}^{\text{fin}}$ in Eq. (\ref{nfin}). Subsequently, we select the signature group size $n$ that satisfies the QDS security condition, as given by Eq. (\ref{nbound}). Finally, the signature rate $R$ is calculated using Eq. (\ref{rate}).

In the optimization process, we search for $(\kappa,\gamma)$ via convex optimization using the CVXR package and $(\mu,x_{\text{th}},p_{\text{sig}},p_{\text{test}})$ via Genetic Algorithm (GA) function in MATLAB, in order to maximize the number of final keys $\hat{N}^{\text{fin}}$ in distribution stage. Subsequently, we try to find the minimal group size $n$ so that it satisfies the QDS security condition. The security parameter $\epsilon_{s}$ is set to be $10^{-10}$. Typical values of the threshold $x_{\text{th}}$ range from 0.4 to 1.5, with a normalization adopted where the vacuum variance is $V=0.25$. With the maximum number of final keys $\hat{N}^{\text{fin}}$ and the minimum group size $n$ after optimization, the best signature rate $R$ can be obtained. More detailed equations for our numerical simulation can be found in Appendix \ref{app_d}.

Figures \ref{compare1} and \ref{compare2} show the simulated signature rates of our protocol in finite-size cases $N=10^{9}-10^{13}$ with excess noise $\xi=0,10^{-3},10^{-2}$. To give a more intuitive presentation of the results, the transmission efficiency $\eta$ is directly replaced by the transmission distance $d$ using an empirical formula of the form $\eta=10^{-0.016d}$ in the context of ultralow loss fiber. In the optimal case, our protocol conduct $10^6$ turns of signing task per second in optimal case and reach a transmission distance of over 40km with considerable signature rate. In Fig. \ref{compare1}, we show the performance of our protocol under different data sizes $N$. Even with a data size as small as $10^{9}$, the signature rate does not drop significantly compared with $10^{13}$ case. In this sense, our protocol is robust against finite-size effects. In Fig. \ref{compare2}, the signature rates of our protocol with different amount of excess noise are presented. It can be seen that with reasonable excess noise $\xi=10^{-2}$, the effective transmission can still reach over a distance of 10km, which demonstrates the feasibility of our protocol in practical implementation with noisy channels.

\begin{table*}[t]
\caption{\label{table1}Comparison of average signature length increasing with message size. The transmission efficiency is set to $0.4$, which is equivalent to 25 km of transmission distance. The data size $N=10^{11}$ and excess noise is $\xi=10^{-3}$.}
\begin{ruledtabular}
\begin{tabular}{cccc}
 Scheme & Signature length $L$ (10b) & Signature length $L$ (1Mb) &  Signature growth ratio\\ \hline
 Ref.~\cite{zhao2021multibit} & $2.50\times 10^6$  &$1.16\times 10^{12}$ & $4.63\times 10^5$ \\
 Ref.~\cite{zhang2019high} & $3.43\times 10^6$ & $1.57\times 10^{11}$ & $4.56 \times 10^4$ \\
 Our scheme & $2.35\times 10^3$ & $2.69\times 10^3$ & $1.14$ \\
\end{tabular}
\end{ruledtabular}
\end{table*}

\subsection{Comparison}
To illustrate the superiority of our proposed protocol, We conduct a comparative analysis with recent CV scheme~\cite{zhao2021multibit}. In existing CV QDS protocols, each message bit is signed independently. When signing multibit messages, an $m$-bit message must be encoded into a longer sequence with length $h$ by inserting ‘0’ or ‘1’ to the original sequence using specific rules. Two efficient QDS encoding rules are employed here~\cite{zhang2019high,zhao2021multibit} (See detailed expressions in Appendix. \ref{app_e}) in our comparison. For these schemes, the signing efficiency, i.e., $\eta_s = m/h$, is obviously less than 1. In contrast, our protocol is natively designed for multibit signing, which means that the encoding process is omitted. Therefore, the efficiency is specifically improved in large-scale tasks. To demonstrate this, our simulations focus on two different scenarios where the message size is $10$ bits (10b) and $10^6$ bits (1Mb), respectively.

Figures \ref{compare3} and \ref{compare4} present the simulated signature length $L$ for signing 10b and 1Mb messages, respectively. The signature length $L$ in the vertical axis refers to the total consumption of signature for a certain message (lower is better), which directly represents the signing efficiency excelling in security and efficiency. In our protocol, the signature length $L$ simply equals the group size $n$. With regard to the encoding rules from Refs. ~\cite{zhang2019high} and ~\cite{zhao2021multibit}, it is calculated by the sum of signature for each message bit. In the horizontal axis, we directly use the transmission efficiency $\eta$ to demonstrate the difference between protocols more clearly. The results shows that our protocol can exhibit a significant advantage on signature length of three to eight orders of magnitude compared with other QDS schemes, marking a quite massive improvement in most practical situations. 

Notably, the comparison of Figs. \ref{compare3} and \ref{compare4} could showcase the stability of our protocol's efficiency across different message size. To accentuate this point, we introduce a new parameter, the signature growth ratio, as an indicator for this property of a QDS protocol. In a nutshell, it is the ratio of signature length for a certain long and short message. In our simulation, we choose the 1Mb and 10b cases respectively. Direct comparison of different protocols is demonstrated in Table \ref{table1}. As the message size grows from 10b to 1Mb, we can find that the signature growth ratio of competing CV QDS protocols are several orders of magnitude larger than ours, which means that the signature length of our protocol grows much slower as message size increases. This observation highlights our protocol's prowess in signing large massages, suggesting its great potential for large-scale practical applications.

\section{\label{sec5}Discussion}
In this paper, we introduce a multibit CV QDS protocol excelling in security and efficiency. Our protocol ensures information-theoretic security against general coherent attacks in the finite-size regime with excess noise in channels. We present a comprehensive security proof for this scenario, employing a cutting-edge fidelity test approach. Furthermore, our protocol incorporates the OTUH signature method with imperfect keys, enhancing the signing efficiency significantly. Numerical simulations reveal that our protocol surpasses previous CV QDS protocols by around eight orders of magnitude in signature rate over a fiber 25 km long, i.e., 0.4 transmission efficiency between the signer and receiver. Importantly, our protocol's signing efficiency remains stable regardless of message size, unlike previous CV QDS protocols, ensuring consistent signature rates even with larger messages, which is advantageous for large-scale commercial applications. In summary, our protocol represents a significant advancement towards the practical realization of multibit CV QDS with information-theoretical security, which should inspire further exploration of the potential of CV QDS in future research endeavors.

Despite the significant contribution of our protocol in terms of security and performance, there are still areas for potential improvement in the future. We briefly discuss some of them just to set the ball rolling.

First of all, while pioneering CV QDS security against general coherent attacks in finite-size regimes, our protocol exhibits a limitation in terms of effective transmission distance. This major drawback is mainly caused by mediocre math conditions in our security analysis, which directly leads to the lack of results below transmission efficiency 0.4. For example, We tend to prioritize simplicity over optimality when calculating the bound satisfying the operator inequality. Additionally, the lack of a better definition for phase error and the inclusion of trash rounds for technical reasons undermine the distribution efficiency. In recent analyses of discrete-modulated CV methods~\cite{lin2019asymptotic,ghorai2019asymptotic}, these methods to achieve longer transmission may provide some help in addressing this existing problem. 

Second, there could be many ways to enhance the signature rate in finite-size regime. A promising route is increasing the number of discrete modulated states beyond two. The fidelity test we use can be straightforwardly generalized to monitoring of such a larger constellation of signals, which would confine the adversary’s attacks more tightly than in the present binary protocol. The feasibility of this idea has been illustrated in Refs.~\cite{lin2019asymptotic,ghorai2019asymptotic}, which use four or more states in signal or test modes and get better performance results of key distribution than the genuine binary methods like ours. Additionally, the composability of OTUH method also offers a rich source of inspiration. The structure of key distribution is not fixed, and any existing or future work in CV QKD, such as Gaussian-modulated CV methods, can be modified for our scheme~\cite{hajomer2024long,zhang2020long,wang2024high,jain2022practical,tian2022experimental,zhang2024continuous,eriksson2020wavelength}. This flexibility opens up numerous possibilities for researches and development in this area.

Furthermore, although set as predetermined value in our simulation, the excess noise and transmission efficiency of the quantum channel need to be accurately estimated in practical implementations. Obviously, a precise estimation of these parameters is crucial for obtaining optimization parameters, which in turn lead to higher signature rates. If we assume that the quantum channel remains consistent over time, a straightforward method for estimation involves conducting several pretest rounds. For instance, if Alice transmits a coherent state with an expectation value of $\sqrt{\mu}$ and a variance of $ 1/4 $, the state received by Bob should have an expectation value of $\sqrt{\eta \mu}$ and a variance of $(1+\xi)/4$, given our assumption of a pure loss channel followed by random displacement. Based on this relationship, Alice and Bob can determine these two parameters by comparing the transmitted and received states. As for specific approaches in different applications, this topic is beyond the scope our current work and will be explored in future experimental research.

Last but not the least, as our security proof techniques are highly scalable, the combination with other effective methods has great prospects in development. For instance, we can generalize our DV inspired approach of estimating the number of phase errors in qubits to the case of qudits (a quantum unit of information that may take any of $d$ states, where $d$ is a variable), which is beneficial for the information density and eventually communication efficiency. Besides, another possible change is trying different hash functions in signing process, e.g., generalized division hash (GDH) in signing process~\cite{shoup1996fast}. A more efficient and secure hash function, which is still compatible with our security proof, could definitely improve the signature performance. 

\begin{acknowledgments}
This work is supported by the National Natural Science Foundation of China (No. 12274223), the Fundamental Research Funds for the Central Universities and the Research Funds of Renmin University of China (No. 24XNKJ14), and the Program for Innovative Talents and Entrepreneurs in Jiangsu (No. JSSCRC2021484).
\end{acknowledgments}

\appendix

\section{\label{app_a}Derivation of the fidelity bound}
To prove the inequality for the fidelity test, we should introduce the theorem below first.
\begin{theo}
Let $\Lambda_{m,r}(\nu) (\nu \geqslant 0)$ be a bounded function given by
\begin{equation}
 	\Lambda_{m, r}(\nu):=e^{-r \nu}(1+r) L_{m}^{(1)}((1+r) \nu) 
\end{equation}
for an integer $m \geqslant 0$ and a real number $r > 0$. Then, we have
\begin{equation}
	\label{EG}
 	\mathbb{E}_{\rho}\left[\Lambda_{m, r}\left(|\hat{\omega}|^{2}\right)\right]=\langle 0|\rho| 0\rangle+\sum_{n=m+1}^{\infty} \frac{\langle n|\rho| n\rangle}{(1+r)^{n}} I_{n, m} 
\end{equation}
where $I_{n,m}$ are constants satisfying $(-1)^m I_{n,m} >0$.
\end{theo}
we see from above equation that, regardless of the value of $r$, the second term on the right side remains zero when $\rho$ has at most $m$ photons. Thus a lower bound on the fidelity between $\rho$ and the vacuum state is given by
\begin{equation}
	\mathbb{E}_{\rho}\left[\Lambda_{m, r}\left(|\hat{\omega}|^{2}\right)\right] \leqslant\langle 0|\rho| 0\rangle \quad(m: odd)
\end{equation}
for any odd integer $m$. Extension to the fidelity to a coherent state $|\beta \rangle$ is straightforward as
\begin{equation}
	\mathbb{E}_{\rho}\left[\Lambda_{m, r}\left(|\hat{\omega}-\beta|^{2}\right)\right] \leqslant \operatorname{Tr}(\rho |\beta\rangle \langle \beta|) \quad(m: odd)
\end{equation}

\section{Details in security analysis}
\subsection{\label{app_b1}Estimation of phase error rate in an n-bit group}
During the examination of unknown information, the phase error rate of each $n$-bit key group $E^{\phi n}$ instead of the phase error rate for all final keys $E^{\phi}$ is required. Fortunately, for a failure probability $\epsilon$, it is easy for us to bound $E^{\phi n}$ from $E^{\phi}$ by using the random sampling without replacement~\cite{yin2020tight}
\begin{equation}
	E^{\phi n} \leqslant E^{\phi} + \gamma^{U}(n,\hat{N}^{\text{fin}}-n,E^{\phi},\epsilon)
	\label{epn}
\end{equation}
where
\begin{equation}
	\gamma^{U}(n, k, \lambda, \epsilon)=\frac{\frac{(1-2 \lambda) A G}{n+k}+\sqrt{\frac{A^{2} G^{2}}{(n+k)^{2}}+4 \lambda(1-\lambda) G}}{2+2 \frac{A^{2} G}{(n+k)^{2}}}
\end{equation}
with $ A=\max \{n, k\} $ and $ G=\frac{n+k}{n k} \ln \frac{n+k}{2 \pi n k \lambda(1-\lambda) \epsilon^{2}} $.

\subsection{\label{app_b2}Solution for the operator inequality}
The aim of this content is to construct $B(\kappa, \gamma)$ which fulfills the operator inequality (\ref{ineq}). Let $\sigma_{\text{sup}}(O)$ denote the supremum of the spectrum of a bounded self-adjoint operator $O$. Although $B(\kappa, \gamma)=\sigma_{\text{sup}}(M[\kappa,\gamma])$ would give the tightest bound satisfying Eq. (\ref{ineq}), it is hard to compute it numerically since system $C$ has an infinite-dimensional Hilbert space. Instead, we would derive a looser but simpler bound in the proposition below.
\begin{prop}
Let $|\beta \rangle$ be a coherent state. Let $\Pi_{\text{ev(od)}}$, $M^{\text{suc}}_{\text{ev(od)}}$, and $M[\kappa,\gamma]$ be as defined in the main text, and define following quantities:	
\begin{equation}
	C_{\mathrm{ev}}:=\left\langle\beta\left|\Pi_{\mathrm{ev}}\right| \beta\right\rangle=e^{-|\beta|^{2}} \cosh |\beta|^{2},
\end{equation}
\begin{equation}
	C_{\mathrm{od}}:=\left\langle\beta\left|\Pi_{\mathrm{od}}\right| \beta\right\rangle=e^{-|\beta|^{2}} \sinh |\beta|^{2},
\end{equation}
\begin{equation}
	\label{dev}
	D_{\mathrm{ev}(\mathrm{od})}:=C_{\mathrm{ev}(\mathrm{od})}^{-1}\left\langle\beta\left|M_{\mathrm{ev}(\mathrm{od})}^{\mathrm{suc}}\right| \beta\right\rangle,
\end{equation}
\begin{equation}
	\label{vev}
	V_{\mathrm{ev}(\mathrm{od})}:=C_{\mathrm{ev}(\mathrm{od})}^{-1}\left\langle\beta\left|\left(M_{\mathrm{ev}(\mathrm{od})}^{\mathrm{suc}}\right)^{2}\right| \beta\right\rangle-D_{\mathrm{ev}(\mathrm{od})}^{2}.
\end{equation}
Let $M^{\mathrm{err}}_{\mathrm{4d}}[\kappa,\gamma]$ and $M^{\mathrm{cor}}_{\mathrm{2d}}[\kappa,\gamma]$ be defined as follows:
\begin{equation}
 	M_{4 \mathrm{d}}^{\mathrm{err}}[\kappa, \gamma]:=\left[\begin{array}{cccc}1 & \sqrt{V_{\mathrm{od}}} & & \\ \sqrt{V_{\mathrm{od}}} & \kappa C_{\mathrm{od}}+D_{\mathrm{od}} & \kappa \sqrt{C_{\mathrm{od}} C_{\mathrm{ev}}} & \\ & \kappa \sqrt{C_{\mathrm{od}} C_{\mathrm{ev}}}, & \kappa C_{\mathrm{ev}}+D_{\mathrm{ev}}-\gamma & \sqrt{V_{\mathrm{ev}}} \\ & & \sqrt{V_{\mathrm{ev}}} & 1-\gamma\end{array}\right] 
\end{equation}
\begin{equation}
 	M_{2 \mathrm{d}}^{\mathrm{cor}}[\kappa, \gamma]:=\left[\begin{array}{cc}\kappa C_{\mathrm{ev}} & \kappa \sqrt{C_{\mathrm{ev}} C_{\mathrm{od}}} \\ \kappa \sqrt{C_{\mathrm{ev}} C_{\mathrm{od}}} & \kappa C_{\mathrm{od}}-\gamma\end{array}\right] 
\end{equation}
Define a convex function
\begin{equation}
 	B(\kappa, \gamma):=\max \left\{\sigma_{\text {sup }}\left(M_{4 \mathrm{d}}^{\mathrm{err}}[\kappa, \gamma]\right), \sigma_{\text {sup }}\left(M_{2 \mathrm{d}}^{\text {cor }}[\kappa, \gamma]\right)\right\} 
\end{equation}
Then, for $\kappa, \gamma \geqslant 0$, we have
\begin{equation}
 	M[\kappa,\gamma] \leqslant B(\kappa, \gamma) \mathbb{I}_{AC} 
\end{equation}
\end{prop}
The detailed proof for this proposition can be found in Ref.~\cite{matsuura2021finite}.

\subsection{\label{app_b3}Functions for finite-size revision}
For the estimation of phase error rate in finite-size scenario, the revision functions $\delta_1,\delta_2$ are required in Eqs. (\ref{delta1}) and (\ref{delta2}). Here we provide the detailed expression of them according to the results of Ref. ~\cite{matsuura2021finite} below.
 
With use of Azuma’s inequality, the function $\delta_1(\epsilon)$ can be obtained as
\begin{equation}
	\delta_{1}(\epsilon):=\left(c_{\max }-c_{\min }\right) \sqrt{\frac{N}{2} \ln \left(\frac{1}{\epsilon}\right)}.
\end{equation}
with $c_{min}$ and $c_{max}$ defined as
\begin{equation}
 	c_{\min } :=\min \left(p_{\mathrm{test}}^{-1} \kappa \min \Lambda_{m, r},-p_{\mathrm{trash}}^{-1} \gamma, 0\right)
\end{equation}
\begin{equation}
	c_{\max } :=\max \left(p_{\mathrm{sig}}^{-1}, p_{\mathrm{test}}^{-1} \kappa \max \Lambda_{m, r}, 0\right)
\end{equation}
The function $\delta_2(\epsilon/2;\hat{N}^{\text{trash}})$ satisfying the bound (\ref{delta2}) on $\hat{Q}_{-}$ can be derived from the fact that $\operatorname{Pr}[\hat{Q}_{-}|\hat{N}^{\text{trash}}]$ is a binomial distribution. The following inequality thus holds for any positive integer $n$ and a real $\delta$ with $0 < \delta <(1-q_{-})n$ (Chernoff bound):
\begin{equation}
 	\operatorname{Pr}\left[\hat{Q}_{-}-q_{-} n \geqslant \delta \mid \hat{N}^{\text {trash }}=n\right] \leqslant 2^{-n D\left(q_{-}+\delta / n \| q_{-}\right)} 
\end{equation}
where
\begin{equation}
 	D(x \| y):=x \log _{2} \frac{x}{y}+(1-x) \log _{2} \frac{1-x}{1-y} 
\end{equation}
is the Kullback-Leibler divergence. On the other hand, for any non-negative integer $n$, we always have
\begin{equation}
	\operatorname{Pr}\left[\hat{Q}_{-}-q_{-} n \leqslant\left(1-q_{-}\right) n \mid \hat{N}^{\text {trash }}=n\right]=1.
\end{equation}
Therefore, for any non-negative integer $n$, by defining $\delta_2(\epsilon; n)$ which satisfies
\begin{equation}
	\begin{aligned}
		D\left(q_{-}+\delta_{2}(\epsilon ; n) / n \| q_{-}\right)&=-\frac{1}{n} \log _{2}(\epsilon)  \left(\epsilon>q_{-}^{n}\right) \\ 
		\delta_{2}(\epsilon ; n)&=\left(1-q_{-}\right) n  \left(\epsilon \leqslant q_{-}^{n}\right).
	\end{aligned}
\end{equation}

\section{\label{app_c}Property of LSFR-based Toeplitz hash}
The property of LSFR-based Toeplitz hash can be concluded as a proposition below.
\begin{prop}
\label{hc}
	For the LFSR-based Toeplitz hash function $h_{p,s}(M)=H_{nm}M$,if $p(x)|M(x)=M_{m-1}x^{m-1}+\ldots+M_1x+M_0$, then $h_{p,s}(M)=0$.
\end{prop}
\begin{prf}
	We define an $n\times n$ matrix $W$ which is only decided by $p$.
	\begin{equation}
 		W=\left(
 			\begin{array}{ccccc}p_{n-1} & p_{n-2} & \ldots & p_{1} & p_{0} \\ 
 				1 & 0 & \ldots & 0 & 0 \\ 
 				0 & 1 & \ldots & 0 & 0 \\ 
 				\ldots & \ldots & \ldots & \ldots & \ldots \\ 
 				0 & 0 & \ldots & 1 & 0
 			\end{array}\right). 
	\end{equation}
	Then we can express $s_i$ in the construction of $H_{nm}$ in Definition \ref{lfsr} through $s$ and $W$
	\begin{equation}
		s_i = W^i s.
	\end{equation}
	Thereafter we can rewrite $h_{p,s}(M)$ as
	\begin{equation}
 		\begin{aligned} 
 			h_{p, s}(M) & =H_{n m} M \\ 
 			& =M_0 s+\sum_{i=1}^{m-1} M_{i} s_i\\ 
 			& =\sum_{i=0}^{m-1} M_{i} W^{i} s \\ 
 			& =M(W) s,
 		\end{aligned} 
	\end{equation}
	where $M(W)=M_{m-1}W^{m-1}+\ldots+m_1 W +m_0 I$ is an $n\times n$ matrix.
	
	Let $f(x)$ be the characteristic polynomial of the matrix $W$, we can calculate it as
	\begin{equation}
 		\begin{aligned} 
 			f(x) & =|x I-W| \\ 
 			& =\left|\begin{array}{ccccc}x+p_{n-1} & p_{n-2} & \ldots & p_{1} & p_{0} \\ 
 			1 & x & \ldots & 0 & 0 \\ 
 			0 & 1 & \ldots & 0 & 0 \\ 
 			\ldots & \ldots & \ldots & \ldots & \ldots \\ 
 			0 & 0 & \ldots & 1 & x\end{array}\right| \\ 
 			& =x^{n}+p_{n-1} x^{n-1}+\ldots+p_{1} x+p_{0} .
 		\end{aligned} 
	\end{equation}
	Obviously, we can find that $f(x) = p(x)$ holds. In other words, $p(x)$ is the characteristic polynomial of the matrix $W$. According to Hamilton-Cayley theorem, we thus have $p(W)=0$. Take a step further, if $p(x)|M(x)$, there is relation $M(W)=0$ and eventually $h_{p,s}(M)=M(W)s=0$.
\end{prf}
In our protocol, we take $p_a$ as $p$ and $Y_a$ as $s$ to construct the hash function in the encryption process. Thus for any generated string $g$ satisfying $p_g(x)|g(x)$, $h(g)=0$ naturally holds according to Proposition \ref{hc} if $p_g=p_a$.

\section{\label{app_d}Models for numerical simulation}
Following the setups of Ref.~\cite{matsuura2021finite}, we normalize quadrature $x$ such that a coherent state $|x\rangle$ has expectation $\langle x\rangle=Re(\omega)$ and variance $\langle(\Delta x)^2\rangle=1/4$.

For the calculation of signature rates, we assume that the communication channel and Bob’s detection apparatus can be modeled by a pure loss channel followed by random displacement. That is to say, the states received by Bob are given by
\begin{equation}
	\rho_{\text {model }}^{(a)}:=\int_{\mathbb{C}} p_{\xi}(\gamma)\left|(-1)^{a} \sqrt{\eta \mu}+\gamma\right\rangle\left\langle(-1)^{a} \sqrt{\eta \mu}+\gamma\right| d^{2} \gamma
\end{equation}
where $\eta$ is the transmission efficiency of the pure loss channel and $p_{\xi}(\gamma)$ is given by
\begin{equation}
	p_{\xi}(\gamma):=\frac{2}{\pi \xi} e^{-2|\gamma|^{2} / \xi}.
\end{equation}
The parameter $\xi$ is the excess noise relative to the vacuum, namely,
\begin{equation}
 	\left\langle(\Delta x)^{2}\right\rangle_{\rho_{\text {model }}^{(\mathrm{a})}}=(1+\xi) / 4.
\end{equation}
We assume that Bob sets $\beta=\sqrt{\eta \mu}$ for the fidelity test. The actual fidelity between Bob’s objective state $|(-1)^a \sqrt{\eta \mu} \rangle$ and the model state $\rho^{(a)}_{\text{model}}$ is given by
\begin{equation}
 	\begin{aligned}
 		&F\left(\rho_{\text {model }}^{(a)},\left|(-1)^{a} \sqrt{\eta \mu}\right\rangle\left\langle(-1)^{a} \sqrt{\eta \mu}\right|\right) \\ 
 		&\quad=\int_{\mathbb{C}} p_{\xi}(\gamma)\left|\left\langle(-1)^{a} \sqrt{\eta \mu} \mid(-1)^{a} \sqrt{\eta \mu}-\gamma\right\rangle\right|^{2} d^{2} \gamma \\ 
 		&\quad=\frac{1}{1+\xi/2}.
 	\end{aligned} 
\end{equation}
For the acceptance probability of Bob’s measurement in the signal rounds, we assume
\begin{equation}
    f_{\text{suc}}(x) = \Theta(|x|- x_{\text{th}}),
    \label{fsuc}
\end{equation}
where $\Theta$ is the standard step function. Thus the acceptance function represents a step function with the threshold $x_{\text{th}} > 0$. In this case, the quantities defined in Eqs. (\ref{dev}) and (\ref{vev}) are given by
\begin{equation}
	\begin{aligned}
		D_{\mathrm{ev}}=&\int_{0}^{\infty} 2 C_{\mathrm{ev}}^{-1} f_{\mathrm{suc}}(x)\left|\left\langle x\left|\Pi_{\mathrm{ev}}\right| \beta\right\rangle\right|^{2} dx \\ 
		= & \frac{1}{4 C_{\mathrm{ev}}}\left[\operatorname{erfc}\left(\sqrt{2}\left(x_{\mathrm{th}}-\beta\right)\right)+\operatorname{erfc}\left(\sqrt{2}\left(x_{\mathrm{th}}+\beta\right)\right)\right. \\ 
		&\left. +2 e^{-2 \beta^{2}} \operatorname{erfc}\left(\sqrt{2} x_{\mathrm{th}}\right)\right],
	\end{aligned}
\end{equation}
\begin{equation}
 	\begin{aligned} 
 		D_{\text {od }}= & \int_{0}^{\infty} 2 C_{\text {od }}^{-1} f_{\text {suc }}(x)\left|\left\langle x\left|\Pi_{\text {od }}\right| \beta\right\rangle\right|^{2} d x \\ 
 		= & \frac{1}{4 C_{\text {od }}}\left[\operatorname{erfc}\left(\sqrt{2}\left(x_{\text {th }}-\beta\right)\right)+\operatorname{erfc}\left(\sqrt{2}\left(x_{\text {th }}+\beta\right)\right)\right. \\ 
 		& \left.-2 e^{-2 \beta^{2}} \operatorname{erfc}\left(\sqrt{2} x_{\text {th }}\right)\right],
 	\end{aligned} 
\end{equation}
\begin{equation}
 	\begin{aligned} 
 		V_{\mathrm{ev}(\mathrm{od})} & =\int_{0}^{\infty} 2 C_{\mathrm{ev}(\mathrm{od})}^{-1}\left(f_{\mathrm{suc}}(x)\right)^{2}\left|\left\langle x\left|\Pi_{\mathrm{ev}(\mathrm{od})}\right| \beta\right\rangle\right|^{2} d x-D_{\mathrm{ev}(\mathrm{od})}^{2} \\ 
 		& =D_{\mathrm{ev}(\mathrm{od})}-D_{\mathrm{ev}(\mathrm{od})}^{2}
 	\end{aligned} 
\end{equation}
where the complementary error function $\operatorname{erfc}(x)$ is defined as
\begin{equation}
 	\operatorname{erfc}(x):=\frac{2}{\sqrt{\pi}} \int_{x}^{\infty} dt \ e^{-t^{2}} 
\end{equation}
We assume that the number of ``succes'' signal rounds $\hat{N}^{\text{suc}}$ is equal to its expectation value
\begin{equation}
 	\begin{aligned} 
 		\mathbb{E}\left[\hat{N}^{\text {suc }}\right] & =\left(\int_{-\infty}^{\infty} f(|x|)\left\langle x\left|\rho_{\text {model }}^{(a)}\right| x\right\rangle d x\right) p_{\text {sig }} N \\ 
 		& =p_{\text {sig }} N\left(P^{+}+P^{-}\right),
 	\end{aligned} 
\end{equation}
where
\begin{equation}
	\label{p}
 	\begin{aligned} 
 		P^{ \pm} & :=\int_{x_{\mathrm{th}}}^{\infty}\langle \pm(-1)^{a} x|\rho_{\text {model }}^{(a)}| \pm(-1)^{a} x\rangle d x \\ 
 		& =\frac{1}{2} \operatorname{erfc}\left(\left(x_{\mathrm{th}} \mp \sqrt{\eta \mu}\right) \sqrt{\frac{2}{1+\xi}}\right).
 	\end{aligned} 
\end{equation}
Similarly, we have the assumption that the number of test rounds $\hat{N}^{\text{test}}$ is equal to $p_{\text{test}}N$ and the number of trash rounds $\hat{N}^{\text{trash}}$ is equal to $p_{\text{trash}}N$. The test outcome $\hat{F}$ is estimated by its expectation value
\begin{equation}
 	\begin{aligned} 
 		\mathbb{E}[\hat{F}] &=p_{\text {test}} N \int_{\mathbb{C}} \frac{d^{2} \omega}{\pi}\langle\omega|\rho_{\text {model }}^{(a)}| \omega\rangle \Lambda_{m, r}\left(\left|\omega-(-1)^{a} \sqrt{\eta \mu}\right|^{2}\right) \\ 
 		& =\frac{p_{\text {test}} N}{1+\xi / 2}\left[1-(-1)^{m+1}\left(\frac{\xi / 2}{1+r(1+\xi / 2)}\right)^{m+1}\right].
 	\end{aligned} 
\end{equation}

Additionally, as for the cost of error correction process $f\hat{N}^{\text{suc}}H(E^{b})$, we set the correction efficiency $f$ to be $1.1$ and the bit error rate $E^{b}$ as
\begin{equation}
 	E^{b}=\frac{P^{-}}{P^{+}+P^{-}}. 
 	\label{eb}
\end{equation}

\section{\label{app_e}Encoding methods for signing a multi-bit message using single-bit QDS}
Suppose the signer needs to sign an $n$-bit message $M=m_{1}\left\|m_{2}\right\| \ldots \| m_{n}, m_{i} \in\{0,1\}, i=1,2, \ldots, n$, where $m_i$ is the $i$-th bit of message and $\|$ represents the concatenation between bits. Also, we denote the signature section with $l$ length for message bit $m_i$ as $s_{m_i}^{l}$.

In Ref.~\cite{zhao2021multibit}, the encoded sequence $\hat{M}$ and its signature $Sig_{\hat{M}}$ is given as
\begin{equation}
	\hat{M}=1\|1\| 1\left\|m_{1}\right\| m_{2}\|\ldots\| m_{n}\|1\| 1 \| 1,
\end{equation}
\begin{equation}
    \begin{aligned}
         Sig_{\hat{M}}= & s_1^{l+c-3}\|s_1^{l+c-2}\|s_1^{l+c-2}\|s_{m_1}^{l+c}\|s_{m_2}^{l+2c}\| \ldots \\
         & \|s_{m_n}^{l+nc}\|s_1^{l+nc+1}\|s_1^{l+nc+2}\|s_1^{l+nc+3}.
    \end{aligned}
\end{equation}
The encoding rule is simply appending several $1$ at the beginning and end of the raw message, which directly gives $h=n+6$. In signing process, we need a growth in signature length with $c (c \geqslant 2)$ intervals to ensure security. 

In Ref.~\cite{zhang2019high}, the encoding rule is to insert a bit $0$ for every $x$ bits in raw message. The length $h$ is calculated to be $h=n+\lfloor\frac{n}{x}\rfloor+2x+4$, where $x$ can be optimized for given message length $n$. When it comes to the signing process, it simply iterates the encoded sequence using single-bit QDS. The expressions are
\begin{equation}
    \begin{aligned}
         \hat{M}= & 1_{1}\|1_{2}\| \ldots\|1_{x+1}\| 0\|m_{1}\| m_{2}\|\ldots\| m_{x} \| 0 \\ 
         & \|m_{x+1}\| m_{x+2}\|\ldots\| m_{2 x}\|0\| \ldots\\
         & \|m_{\lfloor\frac{n}{x}+1\rfloor}\| m_{\lfloor\frac{n}{x}+2\rfloor}\|\ldots\| m_{n}\|0\| 1_{1}\|1_{2}\| \ldots \| 1_{x+1},
    \end{aligned} 
\end{equation}
\begin{equation}
    \begin{aligned}
         Sig_{\hat{M}}= & s_{1_1}^l\|s_{1_2}^l\| \ldots\|s_{1_{x+1}}^l\| s_0^\|s_{m_{1}}^l\|s_{m_{2}}^l\|\ldots\| s_{m_{x}}^l \| s_0^l \\ 
         & \|s_{m_{x+1}}^l\| s_{m_{x+2}}^l\|\ldots\| s_{m_{2 x}}^l\|s_0^l\| \ldots\\
         & \|s_{m_{\lfloor\frac{n}{x}+1\rfloor}}^l\| s_{m_{\lfloor\frac{n}{x}+2\rfloor}}^l\|\ldots\| s_{m_{n}}^l\|s_0^l\| s_{1_{1}}^l\|s_{1_{2}}^l\| \ldots \| s_{1_{x+1}}^l.
    \end{aligned} 
\end{equation}


%

\end{document}